\documentclass[aps,prd,nofootinbib,superscriptaddress,preprintnumbers,amsmath,amssymb,latexsym,array,enumerate,letterpaper,twocolumn]{revtex4}
\usepackage{epsfig}
\usepackage[colorlinks=true,urlcolor=brown,linkcolor=blue,citecolor=magenta]{hyperref}
\usepackage{breakurl}
\usepackage{lineno}
\usepackage{bm,color,xcolor}
\usepackage{slashed} 
\usepackage{graphicx}
\usepackage{soul} 
\usepackage{multirow}
\usepackage{comment}
\usepackage{epstopdf} 
\usepackage{mathtools}
\usepackage{appendix}
\usepackage{url}
\usepackage{doi}
\usepackage[english]{babel}

\makeatletter
\def\simgt{\mathrel{\lower2.5pt\vbox{\lineskip=0pt\baselineskip=0pt
           \hbox{$>$}\hbox{$\sim$}}}}
\def\simlt{\mathrel{\lower2.5pt\vbox{\lineskip=0pt\baselineskip=0pt
           \hbox{$<$}\hbox{$\sim$}}}}
\makeatother

\newcommand{\be}{\begin{equation}}
\newcommand{\ee}{\end{equation}}
\newcommand{\bea}{\begin{eqnarray}}
\newcommand{\eea}{\end{eqnarray}}
\newcommand{\beq}{\begin{eqnarray}}
\newcommand{\eeq}{\end{eqnarray}}

\newcommand{\mPl}{m_{\rm Pl}}

\newcommand{\dd}[1]{\frac{\partial}{\partial #1}}

\newcommand{\dt}{\text{d}}
\newcommand{\BH}{\textrm{BH}}

\def\lsim{\mathrel{\rlap{\lower4pt\hbox{\hskip1pt$\sim$}}
     \raise1pt\hbox{$<$}}}         
\def\gsim{\mathrel{\rlap{\lower4pt\hbox{\hskip1pt$\sim$}}
     \raise1pt\hbox{$>$}}}         

\begin{document}

\widetext

\title{
Correlating Gravitational Wave and Gamma-ray Signals from Primordial Black Holes
}
\author{Kaustubh~Agashe}
\email{kagashe@umd.edu}
\affiliation{Maryland Center for Fundamental Physics, Department of Physics, University of Maryland, College Park, MD 20742, USA}

\author{Jae~Hyeok~Chang} 
\email{jaechang@umd.edu}
\affiliation{Maryland Center for Fundamental Physics, Department of Physics, University of Maryland, College Park, MD 20742, USA}
\affiliation{Department of Physics and Astronomy, Johns Hopkins University, Baltimore, MD 21218, USA}

\author{Steven~J.~Clark} 
\email{steven\_j\_clark@brown.edu}
\affiliation{Department of Physics, Brown University, Providence, RI 02912-1843, USA}
\affiliation{Brown Theoretical Physics Center, Brown University, Providence, RI 02912-1843, USA}

\author{Bhaskar~Dutta} 
\email{dutta@tamu.edu}
\affiliation{Mitchell Institute for Fundamental Physics and Astronomy, Department of Physics and Astronomy, Texas A\&M University, College Station, TX 77845, USA}

\author{Yuhsin~Tsai} 
\email{ytsai3@nd.edu}
\affiliation{Department of Physics, University of Notre Dame, IN 46556, USA}

\author{Tao~Xu} 
\email{tao.xu@mail.huji.ac.il}
\affiliation{Racah Institute of Physics, Hebrew University of Jerusalem, Jerusalem 91904, Israel}

\preprint{
\begin{minipage}{5cm}
\begin{flushright}
UMD-PP-022-01 \\
MI-TH-2112
 \end{flushright}
\end{minipage}
}

\begin{abstract}
Asteroid-mass primordial black holes (PBH) can explain the observed dark matter abundance while being consistent with the current indirect detection constraints. These PBH can produce gamma-ray signals from Hawking radiation that are within the sensitivity of future measurements by the AMEGO and e-ASTROGAM experiments. PBH which give rise to such observable gamma-ray signals have a cosmic origin from large primordial curvature fluctuations. There must then be a companion, stochastic gravitational wave (GW) background produced by the same curvature fluctuations. We demonstrate that the resulting GW signals will be well within the sensitivity of future detectors such as LISA, DECIGO, BBO, and the Einstein Telescope. The multi-messenger signal from the observed gamma-rays and GW will allow a precise measurement of the primordial curvature perturbation that produces the PBH. Indeed, we argue that the resulting correlation between the two types of observations can provide a smoking-gun signal of PBH.
\end{abstract}

\maketitle
\section{Introduction}
Black Holes (BH) are simple but fascinating objects that intertwine gravity and the quantum theory of elementary particles. From the observational side, BH can produce gravitational waves (GW) from merger events, which have been observed by LIGO/VIRGO experiments~\cite{LIGOScientific:2016dsl,LIGOScientific:2020ibl}. Depending on the mass of BH, their Hawking radiation may also produce Standard Model (SM) particles, such as photons, that can be probed via astrophysical observations~\cite{Hawking:1974rv,PhysRevD.13.198}. Since the GW and the electromagnetic (EM) signals are uniquely determined by the BH spin and mass distribution, the observation of both types of the signals can provide valuable information of the BH properties, and the mechanism of the BH production.

Unfortunately, it is not easy for the BH to produce {\em both} strong GW signals from merger events and Hawking radiation that fall in the observable range of the detectors. For example, the observed merger events have BH mass ranges from $m_{\rm BH}=\mathcal{O}(10-100)$ solar mass~\cite{LIGOScientific:2020ibl}, which only corresponds to Hawking temperature $T_H\approx(10^{10}{\rm g}/m_{\rm BH})\,{\rm TeV}=\mathcal{O}(10^{-13}-10^{-12})$~eV and produce photons with frequencies which are way below the experimental sensitivity. The evaporation lifetime $\tau\approx 10^{57}(m_{\rm BH}/M_{\odot})^3$~Gyrs is also too long to produce a sizable signal rate.  It is an interesting question to ask whether there exists a range of BH masses and their associated production mechanism that can generate both GW signals (likely from an {\em alternate} -- to merger -- means) {\em and} Hawking radiation signals, both of which are within the sensitivity of the current or future experiments.

It is easy to determine the range of BH masses that can produce sizable Hawking radiation today. When requiring the lifetime of BH evaporation to be similar to the age of universe -- so the BH still exist and radiate -- we need $m_{\rm BH}\sim 10^{14-15}$~g. The abundance of BH with these masses has been stringently constrained by the measurements of CMB anisotropy~\cite{Acharya:2020jbv}, the extragalactic gamma-ray background~\cite{Carr:2009jm}, and the galactic gamma-ray background~\cite{Carr:2016hva}. Heavier BH can still produce visible signals today as long as their abundance is large enough to compensate the small Hawking radiation rate. If a fraction $f_{\rm BH}$ of dark matter (DM) today is composed of BH with monochromatic mass spectrum, existing observations on the $e^{\pm}$ and neutrinos~\cite{Boudaud:2018hqb,Dasgupta:2019cae}, gamma-ray signals~\cite{Laha:2020ivk,Coogan:2020tuf,DeRocco:2019fjq,Laha:2019ssq,Ray:2021mxu,Ghosh:2021gfa}, radio emission~\cite{Chan:2020zry}, and the heating/cooling of the interstellar medium~\cite{Kim:2020ngi,Laha:2020vhg} have excluded the possible abundance from $f_{\rm BH}\gsim 10^{-8}$ with $m_{\rm BH}=10^{15}$g to $f_{\rm BH}\approx 1$ with $m_{\rm BH}=10^{17}$g: these bounds loosely follow a scaling relation $f^{\rm bound}_{\rm BH}\propto m_{\rm BH}^{4}$ (see~\cite{Carr:2020gox} for a review of the existing bounds). Any possible observations of BH gamma-ray signals need to come from a BH abundance that is below the above exclusion bound, while being above the sensitivity of future experiments.

Next generation detectors, e-ASTROGAM~\cite{e-ASTROGAM:2016bph} and AMEGO~\cite{AMEGO:2019gny}, will explore new territory of gamma-ray signals  between $0.1-100$~MeV with more than one order of magnitude improvement on the signal sensitivity. This opens up new opportunities for seeing the PBH with $f_{\rm BH}$ below the constraints mentioned above. The energy window of the new gamma-ray observation corresponds to monochromatic BH mass around $10^{15}-10^{16}$~g and the experiments can cover $f_{\rm BH}\gsim 10^{-6}(m_{\rm BH}/10^{16}\,{\rm g})^2$~\cite{Coogan:2020tuf}. (see, also~\cite{Ray:2021mxu}.)  Such asteroid-mass BH cannot come from stellar collapse and have to be produced from large density perturbations in the early universe. The idea of such {\em primordial} black holes (PBH) (cf.~from stellar collapse) accounting for the observed DM density has been extensively studied (see, e.g.,~\cite{Carr:2016drx,Sasaki:2018dmp,Green:2020jor,Carr:2020xqk,Carr:2020gox,Villanueva-Domingo:2021spv} for reviews of PBH DM). A cosmological scenario that produces an order one density contrast in the early universe, including models with specific inflaton potentials~\cite{Carr:1975qj,Ivanov:1994pa,Garcia-Bellido:1996mdl,Silk:1986vc,Kawasaki:1997ju,Yokoyama:1995ex,Pi:2017gih,Ashoorioon:2019xqc,Balaji:2022rsy}, first order phase transitions~\cite{PhysRevD.26.2681,Crawford:1982yz,Kodama:1982sf,Moss:1994pi,Freivogel:2007fx,Johnson:2011wt,Kusenko:2020pcg,Baker:2021nyl,Kawana:2021tde,Huang:2022him}, the dynamics of scalar condensates~\cite{Cotner:2016cvr,Cotner:2017tir,Cotner:2018vug,Cotner:2019ykd}, or the collapse of topological defects~\cite{Hawking:1987bn,Polnarev:1988dh,MacGibbon:1997pu,Rubin:2000dq,Rubin:2001yw,Ashoorioon:2020hln,Brandenberger:2021zvn}, provides a means to produce PBH from the gravitational collapse of the perturbation. It is then possible to generate the PBH abundance required for the observable gamma-ray signals.

Remarkably, the large density perturbations in the early universe also source GW \cite{10.1143/PTP.37.831,Ananda:2006af,Baumann:2007zm,Assadullahi:2009jc,Kohri:2018awv,Cai:2019cdl,Domenech:2021ztg}. Once a perturbation mode enters the horizon with an order one density contrast, tensor mode perturbations (hence GW signals) can be produced through the second-order of the density contrasts. Those GW signals can be detected in the current and future GW detectors~\cite{Saito:2008jc,Inomata:2016rbd,Kapadia:2020pir,Kapadia:2020pnr,Sugiyama:2020roc} such as LISA~\cite{LISA:2017pwj,Barausse:2020rsu}, BBO~\cite{Harry:2006fi,Corbin:2005ny}, DECIGO~\cite{Seto:2001qf,Kawamura:2020pcg}, and the Einstein Telescope~\cite{Punturo:2010zz,Maggiore:2019uih}. It is interesting to ask whether for a general assumption of the curvature power spectrum that corresponds to the production of PBH which are visible in gamma-ray searches, can probe the associated GW signals from the same curvature perturbations? Earlier literature (see, e.g.,~\cite{Byrnes:2018txb,Kozaczuk:2021wcl}) has discussed the use of both the GW measurements and the existing bounds on PBH to explore the primordial power spectrum associated with PBH production. However, to the best of our knowledge, there have not been studies on how well the combination of the two types of measurements can determine the PBH properties and confirm that the gamma-ray or GW signals do come from PBH. 

In this work, we show that if we do observe the gamma-ray signals specifically at e-ASTROGAM from PBH produced by collapsing the primordial fluctuations, then the resulting GW signals from the same primordial fluctuations will indeed be visible at future GW detectors. Results produced from AMEGO should be comparable to those from e-ASTROGAM considering they have similar point source sensitivities and angular resolution. We concentrate our analysis on e-ASTROGAM as simply an illustrative choice. Moreover, due to the parametric dependence on the curvature perturbations being different in the fits to the gamma-ray and GW measurements, by {\em correlating} these results we can obtain a smoking gun signal of the asteroid-mass PBH and make a precise measurement of the density fluctuations that seed the PBH.

The outline of this paper is as follows. In the next section, we review the calculation of PBH mass spectrum from a given curvature power spectrum. In Sec.~\ref{sec.GW}, we summarize the calculation of GW from the curvature power spectrum and then calculate the gamma-ray signals from a given PBH mass function in Sec.~\ref{sec.gamma}. In Sec.~\ref{sec.correlate}, we use curvature power spectra with the form of delta-function or log-normal distribution to show that power spectra that produce PBH visible at e-ASTROGAM will generate large signals at future GW detectors. Using three benchmark models of the curvature power spectra, we also estimate the precision of power spectra measurements from e-ASTROGAM, LISA, and BBO and discuss how the combination of these measurements can help identify the PBH and their origin. We conclude in Sec.~\ref{sec.conclusion}.

\section{PBH production from curvature perturbations}\label{sec.Pk}

\begin{figure*}[t]
    \centering
    \includegraphics[width=0.98\columnwidth]{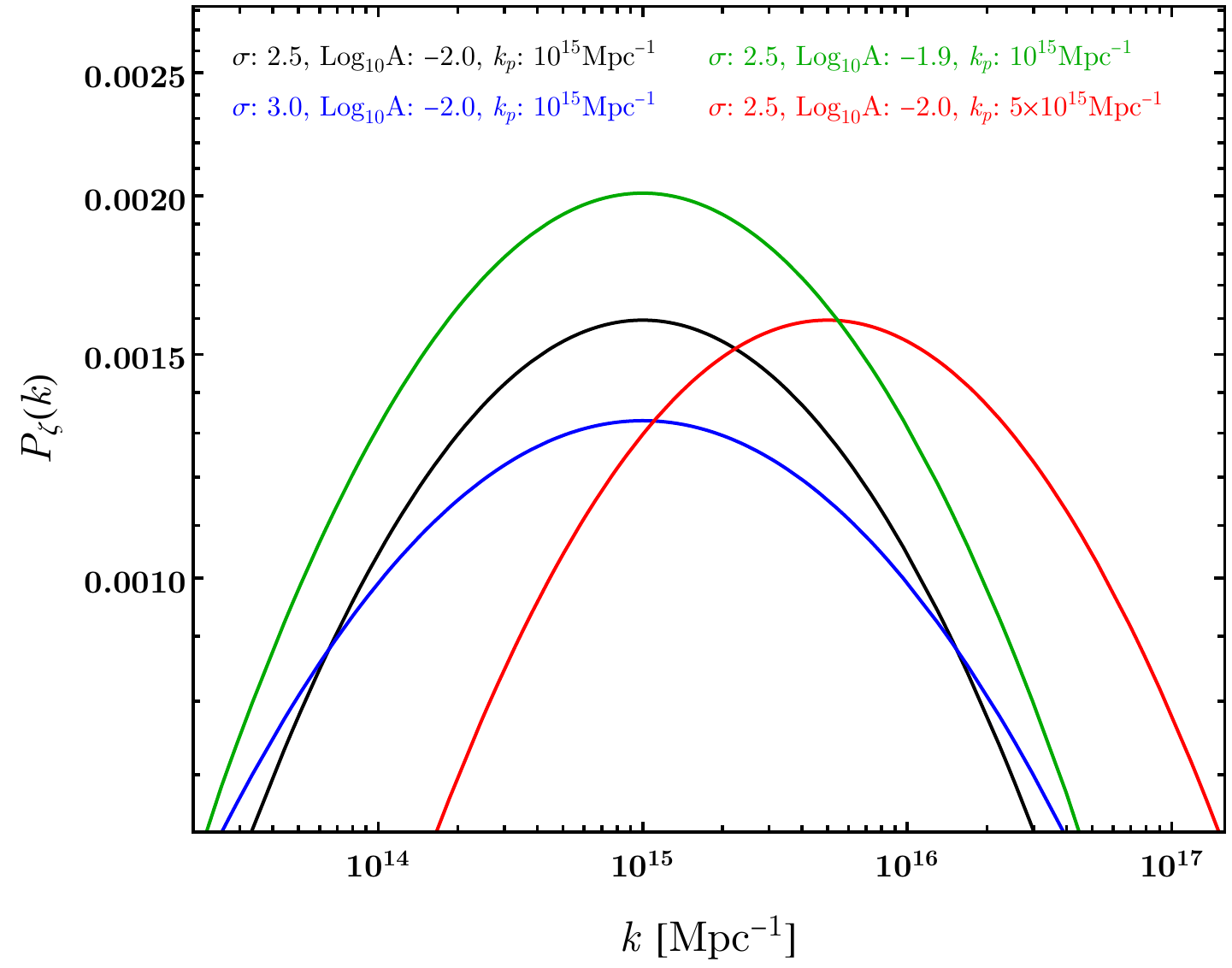}
    $\quad$
    \includegraphics[width=0.98\columnwidth]{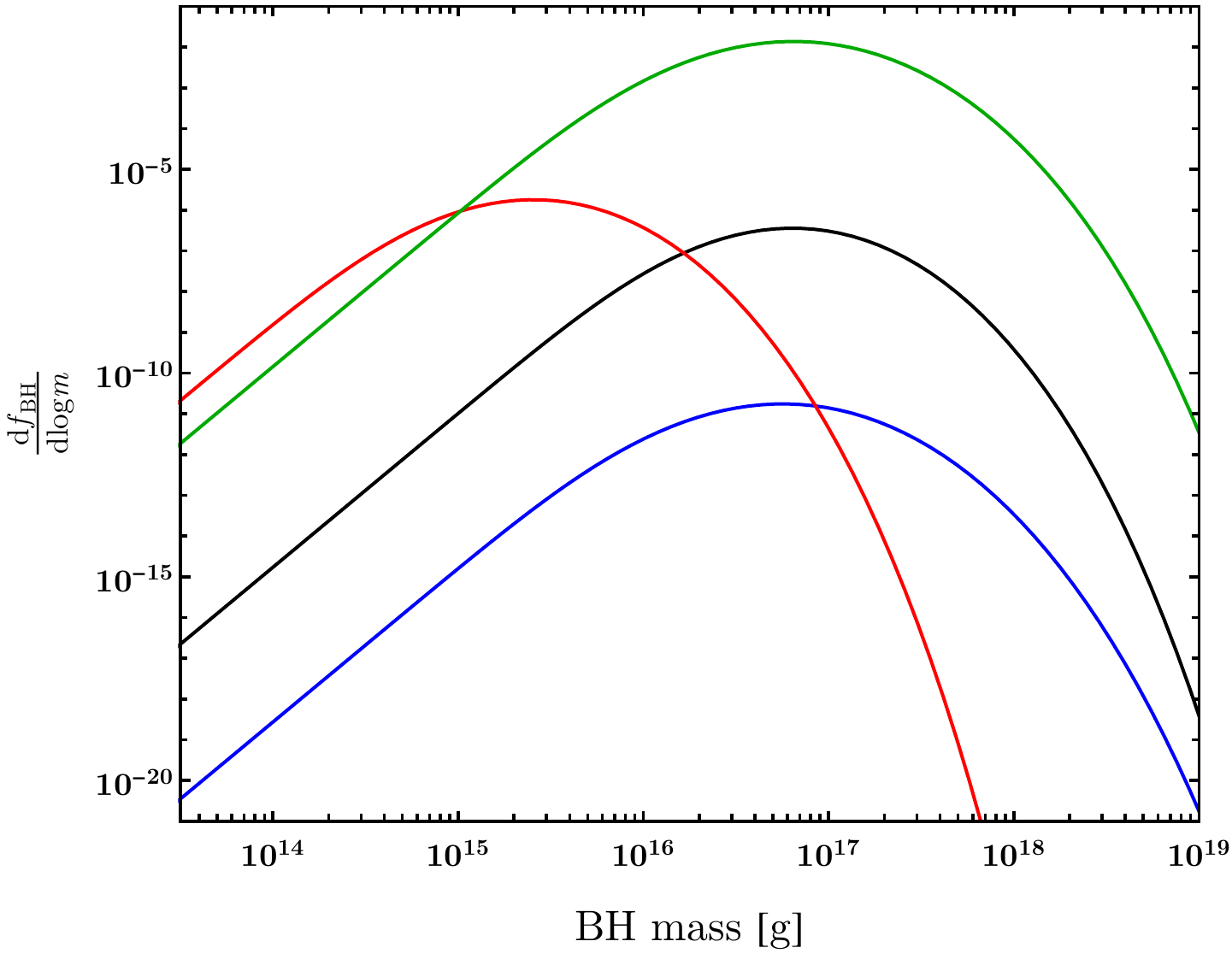}
    \\
    \includegraphics[width=0.98\columnwidth]{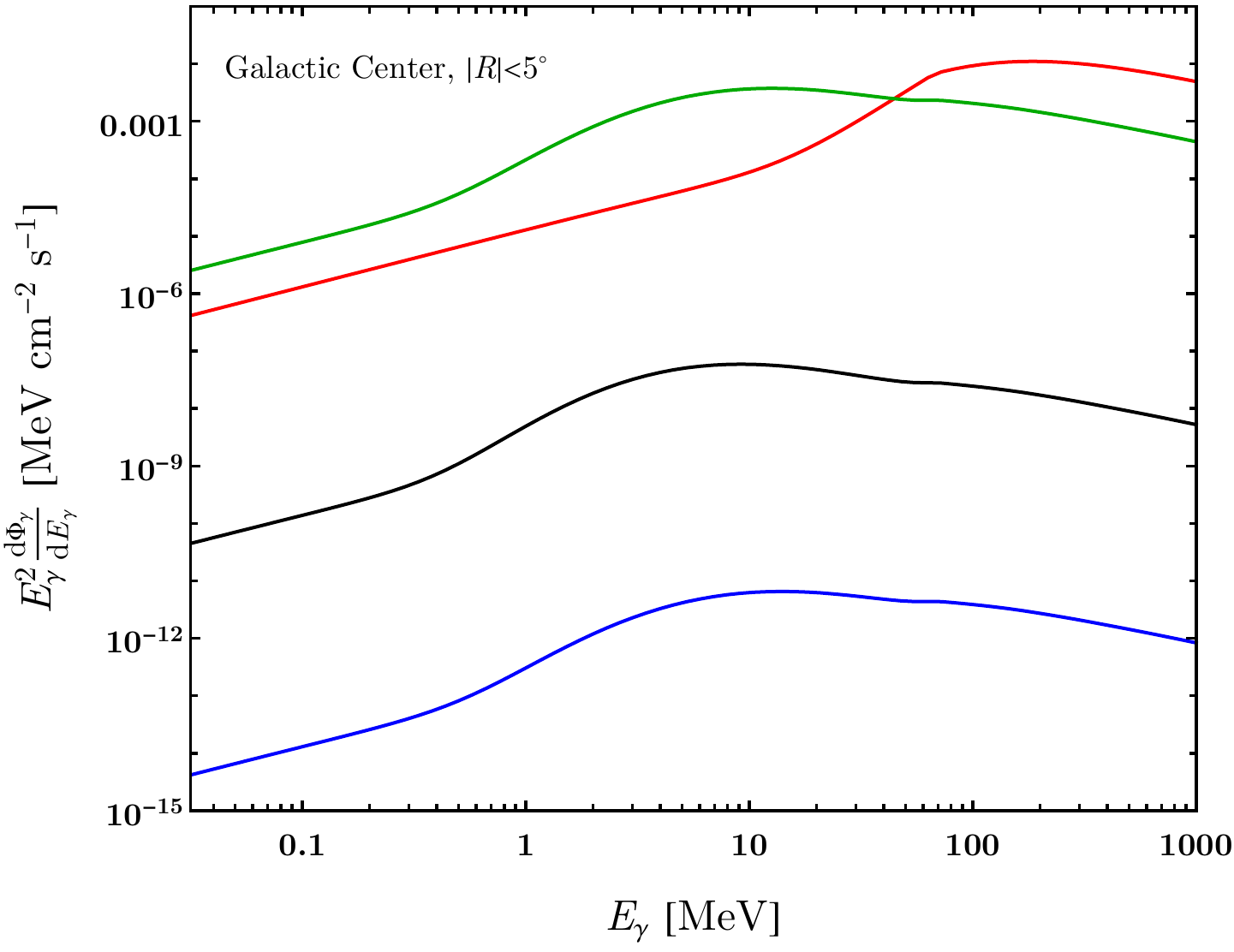}
    $\quad$
    \includegraphics[width=0.98\columnwidth]{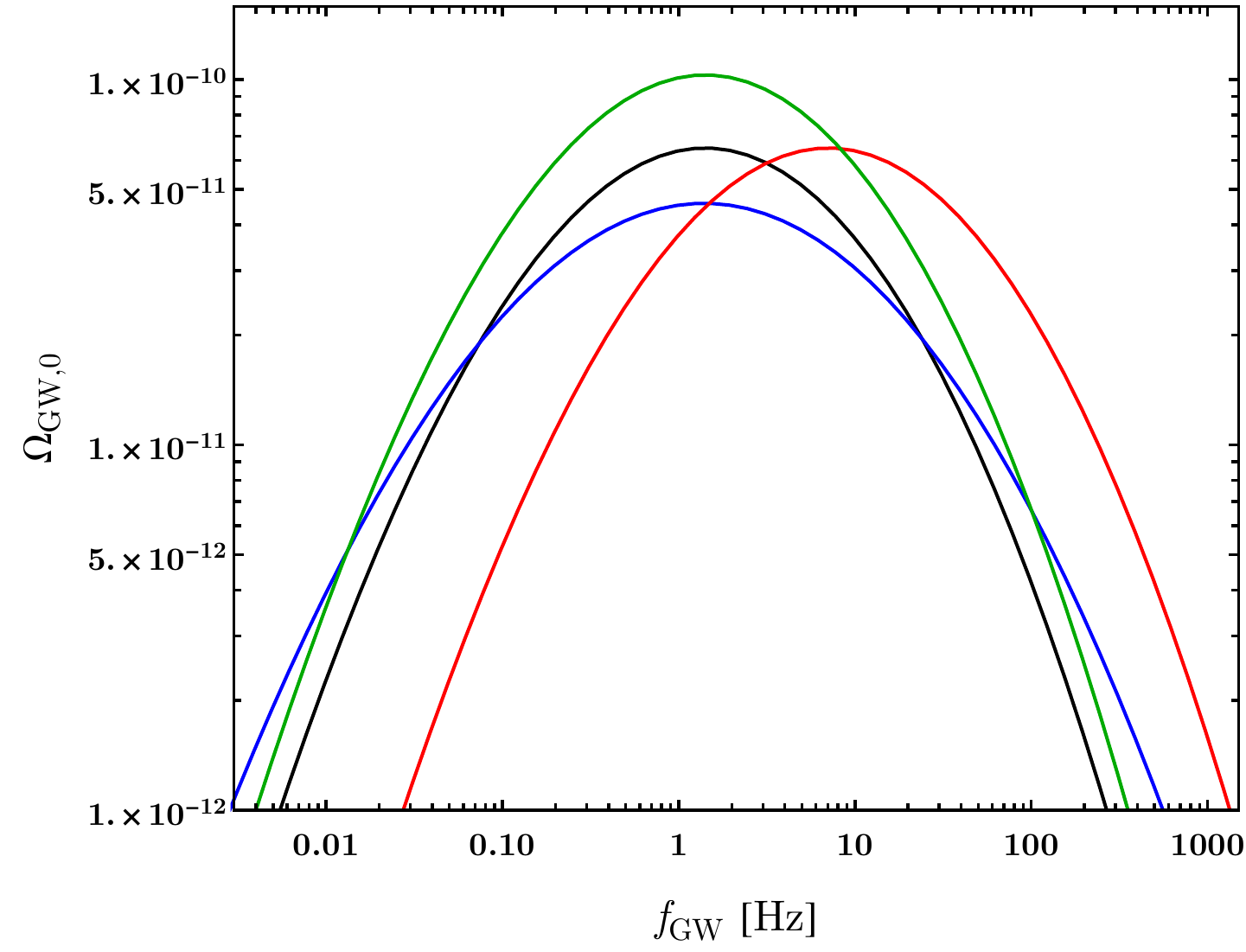}
    \caption{Example plots. Top-left: examples of curvature power spectrum following the log-normal distribution in Eq.~(\ref{eq:pzetaLN}) with different $\sigma$, $A$ and $k_p$ values. Top-right: the resulting PBH mass spectra. The total DM fraction of PBH is $7.1\times10^{-7}$~(black), $3.7\times10^{-11}$~(blue), $3.5\times10^{-6}$~(red) and $3.0\times10^{-2}$~(green) respectively. Bottom-left: the galactic center gamma-ray flux from Hawking radiation in the ROI of $5^{\circ}$. Bottom-right: GW signal spectrum from the curvature power spectrum.} \label{fig:ExamplePLots}
\end{figure*}

To give a concrete example of PBH GW and gamma-ray signals, we consider the large curvature perturbations either takes the form of a monochromatic distribution
\beq
P_{\zeta,\delta}(k)=A_{\delta} \, \delta \left(\log \left(\frac{k}{k_{p,\delta}} \right)\right)\label{eq:pzetaMC}
\eeq
or a log-normal distribution
\beq
P_\zeta(k)=\frac{A}{\sqrt{2\pi\sigma^2}}~\exp\left(-\frac{(\log{k}-\log{k_{p}})^2}{2\sigma^2}\right)\,.
\label{eq:pzetaLN}
\eeq
The power spectrum with a log-normal distribution can be produced, e.g., through the ultra slow-roll (USR) inflation models~\cite{Ballesteros:2017fsr,Dalianis:2018frf,Di:2017ndc} that have been discussed extensively in the context of PBH formation. We take $A$, $k_p$, and $\sigma$ as free parameters when studying the PBH signals. The single field inflation predicts the increase of $P_{\zeta}$ in $k$ to be less than $k^4$~\cite{Byrnes:2018txb}, which corresponds to $\sigma\gsim 1$. When studying gamma-ray and GW signals of an extended power spectrum, we will use $\sigma=2-4$ as examples and pick a range of $(A,k_p)$ that is above the e-ASTROGAM sensitivity while satisfying the existing observational constraints.

The density fluctuations have a chance to collapse into PBH. Since the gamma-ray signal of the PBH is determined by its mass spectrum, the signal estimation is highly sensitive to the estimate of the PBH production from the curvature perturbations. To give a concrete example, we follow the discussion in~\cite{Kozaczuk:2021wcl}, which is based on the Press-Schechter formalism~\cite{1974ApJ...187..425P} with the parameters given in~\cite{Gow:2020bzo} to estimate the PBH mass spectra from $P_\zeta(k)$. Future improvement on the N-body simulation and the understanding of the PBH formation can further tighten the relation between $P_\zeta(k)$ and the mass function.

Consider black holes formed at a particular time corresponding to a horizon size $R$. The probability for the region within the horizon to carry a smoothed density contrast $\delta$ can be described by
\beq
p(\delta)=\frac{1}{\sqrt{2\pi\sigma^2_0}}e^{-\frac{\delta^2}{2\sigma_0^2}}\,.
\label{eq:prob}
\eeq
The variance of density contrasts on the scale $R$ is given by
\beq
\sigma_0^2=\displaystyle{\int_0^{\infty}}\frac{{\rm d}k}{k}\frac{16}{81}(kR)^4W^2(k,R)P_\zeta(k),
\label{eq:sigma0}
\eeq
where $W(k,R)$ is the window function that smooths the power spectrum with scale $k$ on the horizon size $R$. In our analysis we take
\beq
W(k,R)=\exp\left[-\frac{(kR)^2}{4}\right]\,.
\label{eq:window}
\eeq
In Press-Schechter, we assume perturbation modes with density contrasts larger than a threshold $\delta_c$ collapse into black holes. Note that different choices of window functions can lead to quite different PBH abundance from the same $P_{\zeta}$~\cite{Ando:2018qdb}, but~\cite{Young:2019osy,Gow:2020bzo} shows that the deviations should be within $10\%$ if one uses consistent window functions and threshold density contrasts.

If defining $\beta_R$ as the fraction of energy density in the BH that formed at the time with horizon size $R$  
\beq
\beta_R\equiv\frac{\rho_{{\rm BH},R}}{\rho_{r,R}}\,,
\label{eq:betaR}
\eeq
where $\rho_r$ is the radiation energy density, we can calculate $\beta_R$ contributed by $\delta$ of a given horizon as
\beq
\frac{{\rm d}\beta_R}{{\rm d}\delta}=\frac{2m(R,\delta)}{M_H(R)}\,p(\delta)\,,\quad\delta\geq\delta_c\,.
\label{eq:betaRint}
\eeq
Here $M_H$ is the average horizon mass with size $R$, and the BH mass produced by a given density contrast follows the relation~\cite{Choptuik:1992jv,Niemeyer:1997mt, Niemeyer:1999ak}
\beq
m(R,\delta)=M_H(R)K(\delta-\delta_c)^{\gamma}\,.
\label{eq:m}
\eeq
The expression is valid with $(\delta-\delta_c)\lsim 10^{-2}$~\cite{Musco:2020jjb} and holds down to $(\delta-\delta_c)\sim10^{-10}$~\cite{Musco:2008hv, Musco:2012au}. The dimensionless $K$, $\delta_c$, and $\gamma$ are constants determined via numerical simulations. In this work we use $K=10$, $\delta_c=0.25$, and $\gamma=0.36$ given in~\cite{Young:2020xmk}, which come from translating the results in~\cite{Young:2019yug,Niemeyer:1997mt,Musco:2008hv,Musco:2012au} with a top-hat smooth function to the result with a Gaussian smooth function Eq.~(\ref{eq:window}).

The total energy density fraction in PBH is obtained by integrating Eq.~(\ref{eq:betaRint}) from all possible horizon sizes. Changing the variable from $\delta$ to $m$ using Eq.~(\ref{eq:m}) and including the redshift factor of the PBH density from the production time, we can obtain the energy fraction of PBH at matter-radiation equality as a function of PBH mass 
\beq
\frac{{\rm d}\beta_{\rm eq}}{{\rm d}m}=\displaystyle{\int_0^{\infty}} \frac{{\rm d}R}{R}\frac{R_{\rm eq}}{R}\,\left(2 \frac{g_{\star s,\textrm{eq}}^{4/3}}{g_{\star s,R}^{4/3}}\frac{g_{\star,R}}{g_{\star,\textrm{eq}}} \right)^{1/2} \frac{{\rm d}\beta}{{\rm d}m}\,.
\label{eq:betam}
\eeq
Here $g_{\star}$ and $g_{\star s}$ are the number of relativistic degrees of freedom for energy density and entropy, and subscript $\textrm{eq}$ is for the time at matter-radiation equality. The relevant PBH mass for the indirect detection searches comes from the collapse of fluctuations with  $k_p\sim10^{14-15}~{\rm Mpc}^{-1}$. These perturbation modes enter the horizon at $z\sim 10^{20-21}$, and the PBH formation happens deep inside the radiation dominated era. Thus, we use $g_{\star,R}=g_{\star s,R}=106.75$ in this calculation. For matter-radiation equality, we take the value $g_{\star,{\rm eq}}=3.36$ and $g_{\star s,{\rm eq}}=3.91$~\cite{Kolb:1990vq}.

Now we define the mass function at matter-radiation equality as
\beq
\frac{{\rm d}f_{\rm BH,eq}}{{\rm d}m} =  \frac{\textrm{d}}{\textrm{d}m} \left( \frac{\rho_\textrm{BH,eq}}{\rho_\textrm{CDM,eq}} \right) = \frac{\Omega_m}{\Omega_{\rm CDM}} \frac{{\rm d}\beta_{\rm eq}}{{\rm d}m} \,,
\eeq
where $\Omega_m$ and $\Omega_{\rm CDM}$ are the energy fraction of matter and cold dark matter today. 

The mass function can be fitted to a log-normal distribution
\bea
\frac{{\rm d}f_{\rm BH}}{{\rm d}m} = \frac{f_{\rm tot}}{m\sqrt{2\pi}\sigma_m}\exp\left(-\frac{\log^2(m/m^{\rm peak})}{2\sigma^2_m}\right).
\label{eq:lognormalfit}
\eea
The location of the mass function peak can be expressed with the horizon mass corresponding to the power spectra peak as 
\bea
m^{\rm peak}&=&\gamma_{\rm eff} M_{H}(R=k^{-1}_{p})\label{eq.mpeak}\\
&\simeq&2\times 10^{16}~{\rm g}\times\gamma_{\rm eff}~\left(\frac{k_{p}}{10^{15}~{\rm Mpc}^{-1}}\right)^{-2}\,,\nonumber
\eea
where the $\gamma_{\rm eff}\sim\mathcal{O}(1)$ factor is studied in detail in \cite{Kozaczuk:2021wcl}.

Finally, we consider the mass loss from Hawking radiations to get the actual PBH mass function today, $f_\textrm{BH}$. Using an approximation of a constant degree of freedom of Hawking radiation, we get
\bea
\frac{{\rm d}f_{\rm BH}}{{\rm d}m}&=& \frac{m^3}{m^3+3 f_0 \mPl^4 t_0} \left. \frac{{\rm d}f_{\rm BH,eq}}{{\rm d}m'} \right|_{\tiny m'=(m^3+3 f_0 \mPl^4 t_0)^{1/3}}\,. \nonumber\\
\label{eq:fm}
\eea
Here $t_0$ is the age of the universe, and $f_0 \approx 1.895\times 10^{-3}$. Again, we assume PBH formation happens at deep inside the radiation domination, so the age of PBH is $t_0$. Please see Appendix.~\ref{App:massfunction} for details. 

In Fig.~\ref{fig:ExamplePLots}, we show some examples of the log-normal curvature power spectra (upper-left) and the resulting PBH mass function (upper-right). The mass distribution is very sensitive to the value of $(A,\sigma)$. Since the gamma-ray signal is determined by the mass function, an observation of the signal can provide a precise measurement of the power spectrum. Curvature fluctuations with higher $k$-modes produce PBH formation with a smaller horizon size. This leads to a mass function peaked at a smaller $m$ (red vs. black). The PBH formation rate is sensitive to three parameters in Eq.~(\ref{eq:m}), especially $\delta_c$. The precise value of $\delta_c$ should depend on the shape of the curvature power spectrum, such as the width and non-Gaussianity ~\cite{Franciolini:2018vbk,Musco:2020jjb}. In Appendix~\ref{app.deltac}, we calculate the expected signal using different values of $\delta_c$ to illustrate the validity of our results, even if the precise value of $\delta_c$ is a factor of a few different from $0.25$. Because the gamma-ray signal is much more sensitive to $A$ than the GW signal, as long as e-ASTROGAM sees the gamma-ray signal, the corresponding $P_{\zeta}(k)$ (with a slightly different $A$ in the new estimate) will produce GW energies within an order of magnitude of our estimate. The uncertainty in $\delta_c$ therefore does not change our statement on the visibility of both the gamma-ray and GW signals from the PBH model.  

The large curvature perturbations can produce cosmological signals other than the PBHs, for example, spectral distortions that can be used to constrain $P_\zeta$ with a much lower $k\lsim 10^5$~Mpc$^{-1}$~\cite{Kohri:2014lza,Nakama:2017xvq}. Another signal produced by the perturbation is a stochastic GW background which we will discuss in the next section.

\section{GW signals from the curvature perturbation}\label{sec.GW}

As mentioned before, large curvature perturbations produce a stochastic GW background. Here we briefly review the formalism to calculate GW induced at second order in curvature perturbations. We consider the GW production during the radiation-dominated era and follow the calculation in~\cite{Kohri:2018awv}, taking the conformal Newtonian gauge when expressing the primordial scalar perturbation $\Phi$. The scalar perturbation in the quadratic order provides a source of the tensor mode perturbations $h$. The tensor and the curvature power spectra are related to the tensor and scalar perturbations by
\begin{eqnarray}
P_h(\eta,k)\delta^3({\bf k+k'})&=&\frac{k^3}{2\pi^2}\langle h_{\bf k}(\eta)h_{\bf k'}(\eta)\rangle\,, \\ 
P_{\zeta}(\eta,k)\delta^3({\bf k+k'})&=&\frac{k^3}{2\pi^2}\langle \Phi_{\bf k}(\eta)\Phi_{\bf k'}(\eta)\rangle\,.\label{eq.Ph}
\end{eqnarray}
The energy density of the GW for a given mode $k$ at a conformal time $\eta$ is 
\begin{equation}
\Omega_{\rm GW}(\eta,k)=\frac{1}{24}\left(\frac{k}{a(\eta)H(\eta)}\right)^2P_h(\eta,k)\,.\label{eq.GW}
\end{equation}
In the radiation-dominated era, the dimension-less GW power spectrum is given by
\bea
P_h(\eta,k)&\simeq&2\int^{\infty}_{0}dt\int^{1}_{-1}ds\left(\frac{t(t+2)(s^2-1)}{(t+s+1)(t-s+1)}\right)^2\nonumber\\
&&\times I^2(s,t,k\eta)P_\zeta(u k)P_\zeta(v k)\,,
\eea
where $P_\zeta$ is the primordial curvature perturbation,  $u=\frac{t+s+1}{2}$ and $v=\frac{t-s+1}{2}$. The $I^2$ term contains an integral over some combination of the Green's function for the tensor modes and the transfer functions for the scalar modes. Since curvature perturbations decay quickly after horizon re-entry in the radiation-dominated era, GW signals are mainly produced right after re-entry. When calculating the GW energy, we include GW with $k\gg \eta^{-1}$ so that the wave is subhorizon and has a well-defined energy density. In this case, $I^2$ can be approximated as  
\begin{widetext}
\bea
I^2(s,t,k\eta)&=&\frac{288(s^2+t(t+2)-5)^2}{k^2\eta^2(t+s+1)^6(t-s+1)^6}\bigg(\frac{\pi^2}{4}\Big(s^2+t(t+2)-5)^2\Theta(t-(\sqrt{3}-1)\Big)\nonumber\\
&&\qquad\qquad+\Big(-(t+s+1)(t-s+1)+\frac{1}{2}(s^2+t(t+2)-5)\log\Big|\frac{t(t+2)-2}{3-s^2}\Big|\Big)^2\bigg)
\eea
\end{widetext}
After the GW energy density is calculated at the re-entry time $\eta_c$, we evolve it to the current time as GW evolves as radiation
\beq
\Omega_{\rm GW}(\eta_0,k)=0.83\left(\frac{g_{\star,c}}{10.75}\right)^{-\frac{1}{3}}\Omega_{r,0}\Omega_{\rm GW}(\eta_c,k).
\eeq
The energy density of radiation today is $\Omega_{r,0}=8.5\times10^{-5}$. The effective massless degree of freedom at reentry time $g_{\star,c}$ is calculated through solving 
\beq
\frac{k}{k_{\rm eq}}=2(\sqrt{2}-1)\left(\frac{g_{\star s,\rm eq}}{g_{\star s}(T)}\right)^{\frac{1}{3}}\left(\frac{g_{\star}(T)}{g_{\star,\rm eq}}\right)^{\frac{1}{2}}\frac{T}{T_{\rm eq}}
\eeq
We used the measured value $T_{\rm eq}=0.8~{\rm eV}$.

\section{Gamma-ray signals from the asteroid-mass PBH}\label{sec.gamma}

\begin{figure*}[t]
    \centering
    \includegraphics[width=0.98\columnwidth]{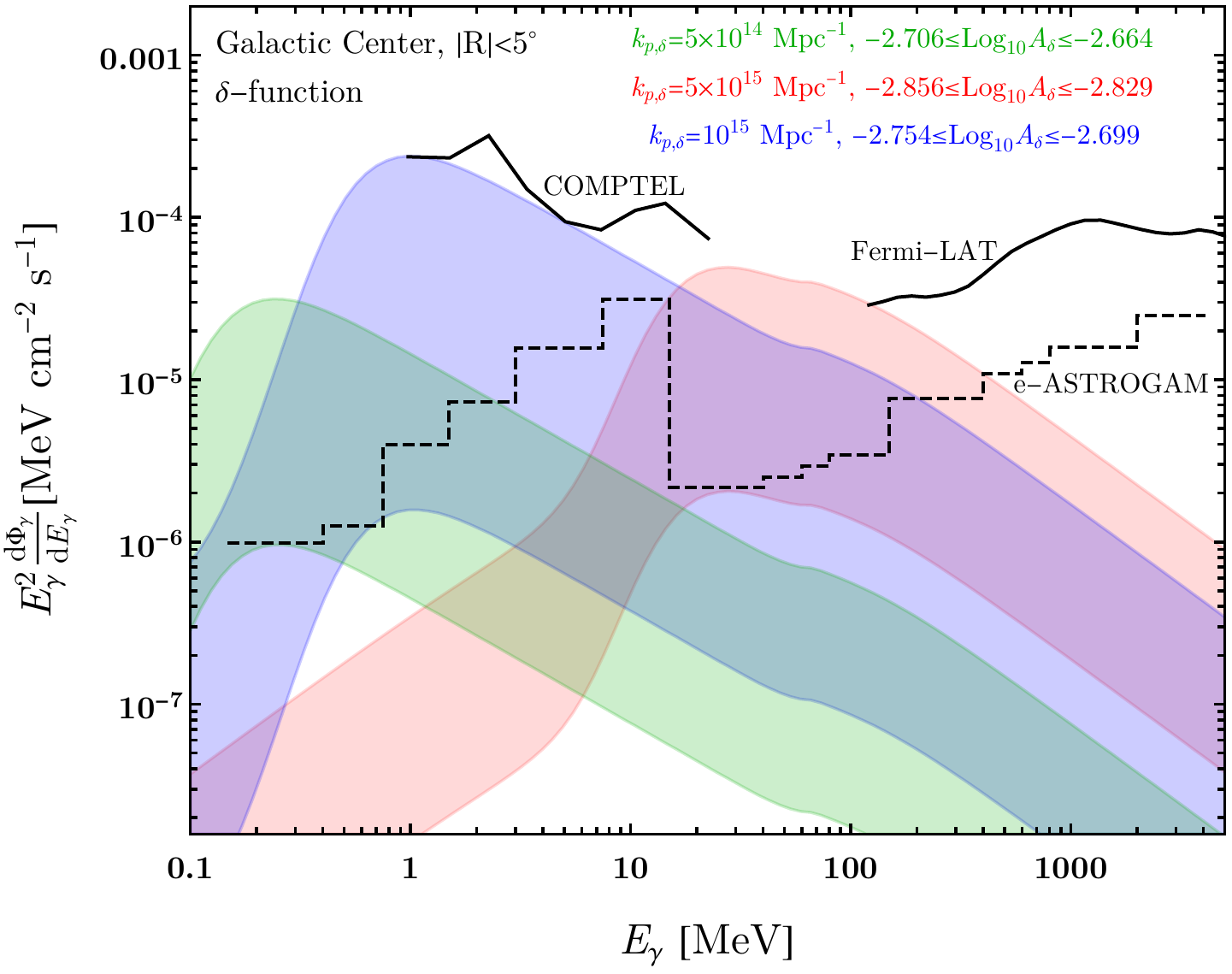}
    $\quad$
    \includegraphics[width=0.98\columnwidth]{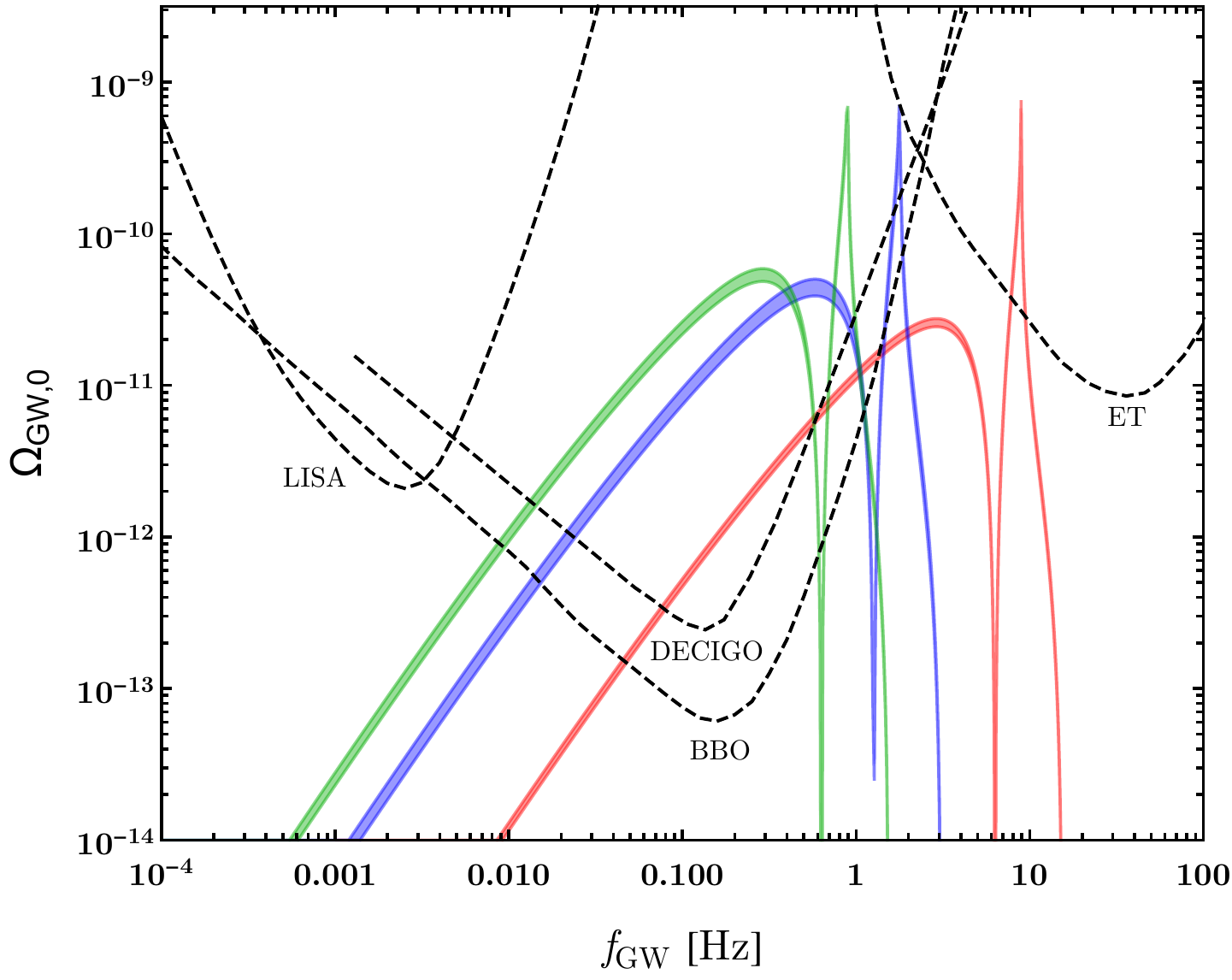}
    \\
    \includegraphics[width=0.98\columnwidth]{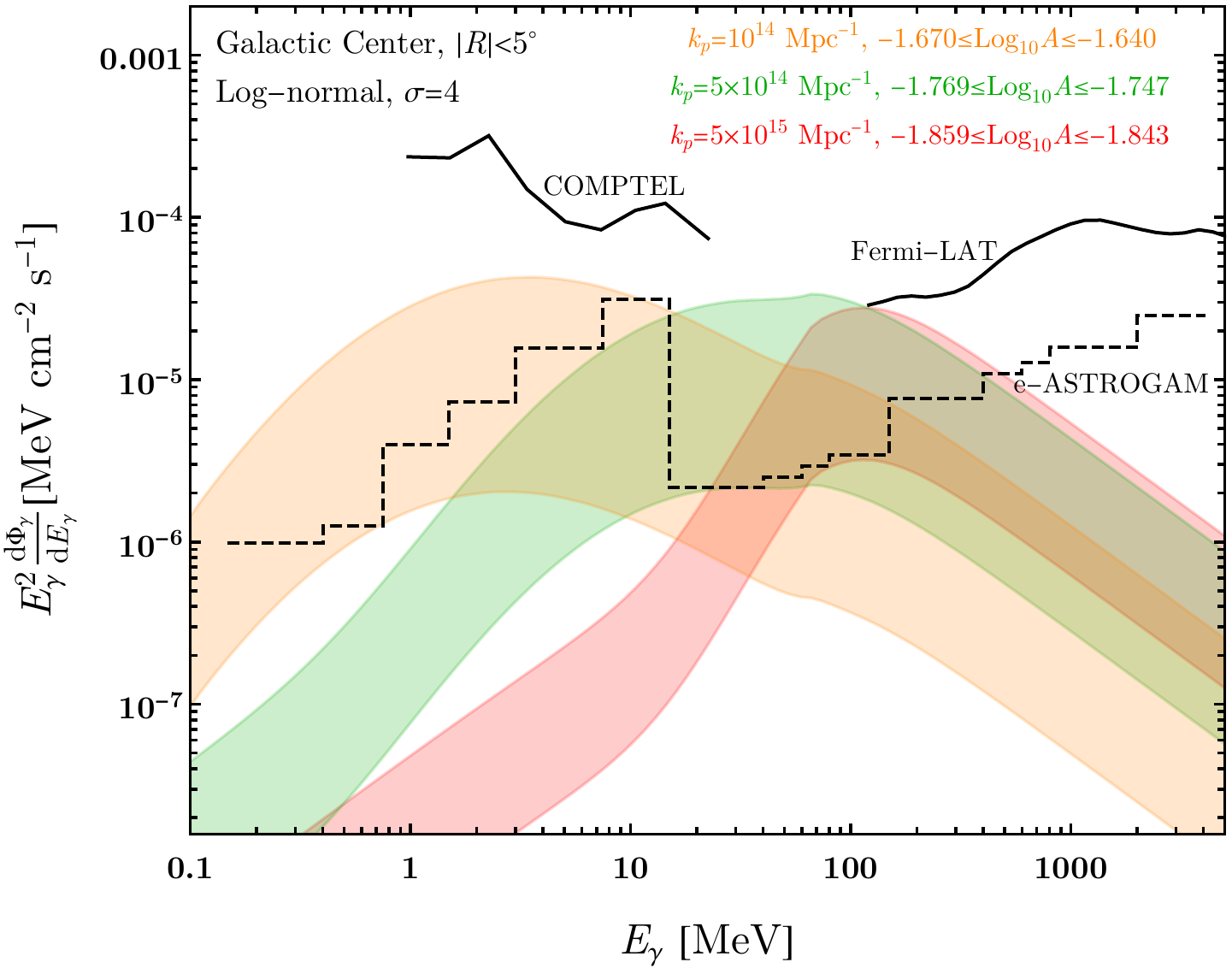}
    $\quad$
    \includegraphics[width=0.98\columnwidth]{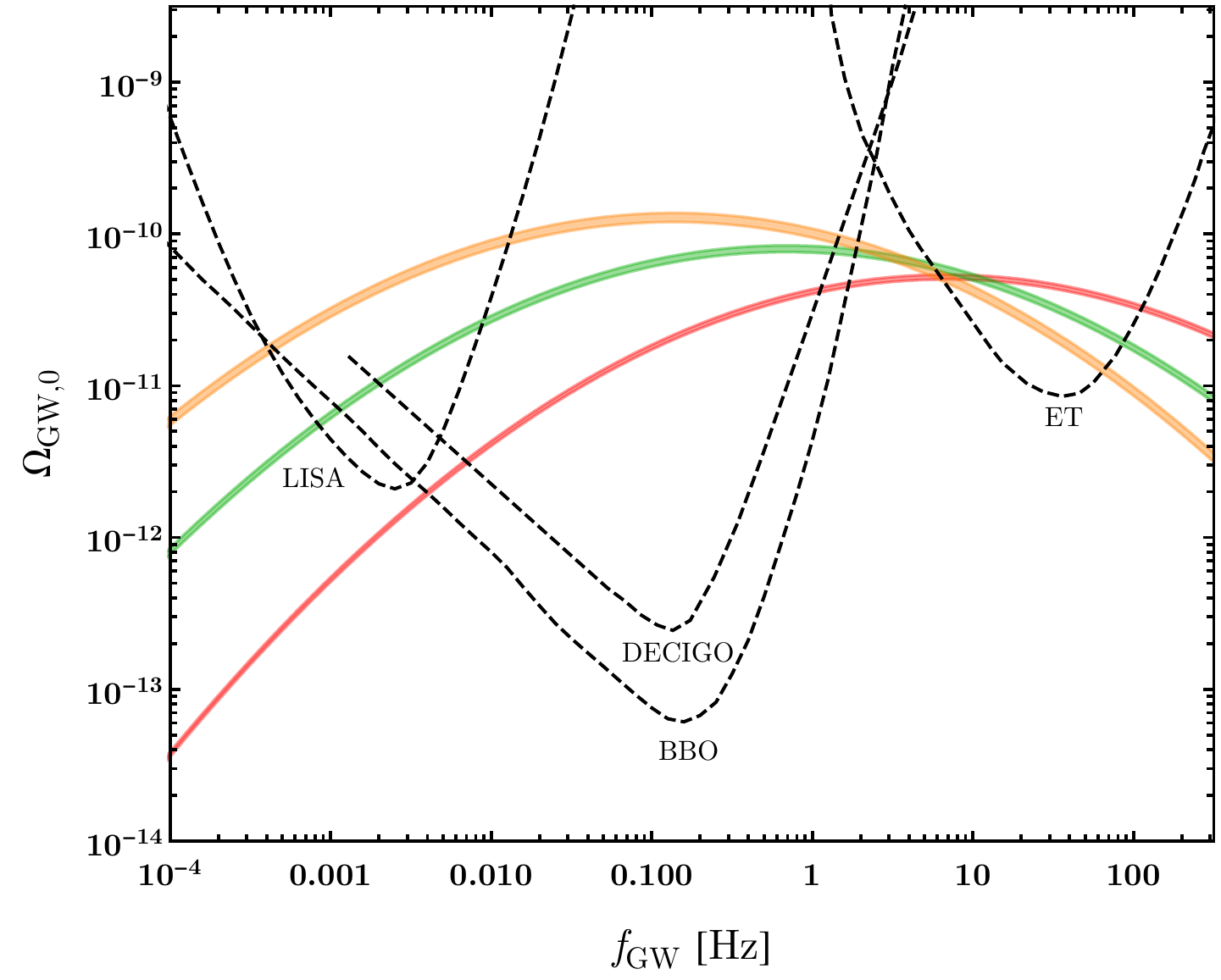}
    \caption{Range of gamma-ray and GW signals from $\delta$-function power spectrum (top row) and log-normal power spectrum of $\sigma=4$ (bottom row). We choose three $k_{p,\delta}/k_{p}$ values that cover the e-ASTROGAM sensitivity region. In the gamma-ray plots, the upper edges of the bands are determined when the gamma-ray flux just touch the Fermi-LAT(rescaled) bound, COMPTEL (rescaled) bound or when $f_{\rm BH, total}=1$, which sets the upper boundaries of the green ($\delta$-function) and orange bands. The lower edge of the bands are determined when the gamma-ray flux is just above the future e-Astrogam sensitivity curve. The e-ASTROGAM sensitivities for each bin are at the $3\sigma$ level over the background and LISA and BBO are at the $1\sigma$ level. The DECIGO and ET sensitivity curves are taken from~\cite{Kawamura:2020pcg,Maggiore:2019uih}.} \label{fig:SignalBand}
\end{figure*}

After being produced from the collapse of density fluctuations, PBH lose their mass through Hawking radiation. Especially, PBH with mass $\gsim 10^{15}\,$g have lifetimes longer than the age of universe and are still emitting gamma-rays today, which can be detected in future observations. We assume only Schwarzschild PBH in this work with negligible spin for correcting the radiation signals. This is a reasonable assumption for the PBH formed from the spherical collapse of large primordial density fluctuations that reenter the horizon during the radiation-dominated era~\cite{Chiba:2017rvs, Mirbabayi:2019uph,DeLuca:2019buf}. In this case, the Hawking temperature of PBH of mass $m$ is
\beq
T_{\rm BH}=\frac{1}{8\pi G_{N} m}=1.05\left(\frac{10^{16}\,{\rm g}}{m}\right)\,{\rm MeV}\,,
\eeq
where $G_{N}=\mPl^{-2}$ is the Newton's constant. For a PBH with mass $m$, each particle species $i$ with mass $m_i$ and degree of freedom $g_i$ is emitted with energy $E_i$ at a rate
\begin{equation}
    \frac{\partial N_{i,\textrm{primary}}}{\partial E_i \partial t} = \frac{g_i}{2 \pi}\frac{\Gamma_i(E_i,m,m_i)}{e^{E_i/T_{\rm BH}}\pm 1}\,.
\label{eq:HRprimary}
\end{equation} 
where $+/-$ is for fermion/boson. The greybody factor $\Gamma_i=\frac{\sigma_i}{\pi}(E_i^2-m_i^2)$ accounts for the absorption probability at the horizon. When $E_i\gg T_{\rm BH}$, $\sigma_i=27\pi G_N^2 m^2$ is the geometrical optics limit. $\sigma$ decreases when $E_i$ is small, and it also depends on the spin of the emitted particle. In this study, we take the greybody factor calculated in BlackHawk~\cite{Arbey:2019mbc, Arbey:2021mbl}.

The contribution to gamma-ray flux consists of both primary photons and secondary photons from PBH Hawking radiation. The primary photon flux is directly calculated with Eq.~(\ref{eq:HRprimary}). The secondary photon flux is from the final state radiation~(FSR) of primary charged particles and the decay of primary neutral pions into photons. The total photon flux from a PBH is expressed as
\begin{eqnarray}\label{eq:gammaraytot}
    \frac{\partial N_{\gamma,\textrm{tot}}}{\partial E_\gamma \partial t} &=& \frac{\partial N_{\gamma,\textrm{primary}}}{\partial E_\gamma \partial t} \nonumber\\
    &&+ \sum_{i=e^\pm,\mu^\pm,\pi^\pm} \int d E_i \frac{\partial N_{i,\textrm{primary}}}{\partial E_i \partial t} \frac{d N_{i,\textrm{FSR}}}{dE_\gamma} \nonumber\\
    &&+ \sum_{i=\pi^0} \int d E_i 2\frac{\partial N_{i,\textrm{primary}}}{\partial E_i \partial t} \frac{d N_{i,\textrm{decay}}}{dE_\gamma}\,,
    \label{eq:totflux}
\end{eqnarray}

The second term is the FSR rate. Primary charged particles with energy $E_i$ can radiate photons during their production. The universal FSR energy spectrum at leading order in $\alpha$ can be calculated as 
\begin{eqnarray}
    \frac{d N_{i,\textrm{FSR}}}{dE_\gamma} &=& \frac{\alpha}{\pi Q_i}P_{i\rightarrow i\gamma}(x) \left [\log \left (\frac{1-x}{\mu_i^2} \right ) -1 \right ]\label{eq.FSR1}\\
    P_{i\rightarrow i\gamma}(x) &=& \begin{dcases} \frac{2(1-x)}{x}, & i=\pi^\pm \\ \frac{1+(1-x)^2}{x}, & i=\mu^\pm, e^\pm\label{eq.FSR2} \end{dcases}
\end{eqnarray}
where $x=2E_\gamma/Q_i$, $\mu_i=m_i/Q_i$ and we choose the FSR energy scale $Q_i=2E_i$. The splitting function $P_{i\rightarrow i\gamma}(x)$  distinguishes between bosons and fermions. The FSR rate is large for small $E_\gamma$, so it determines the shape of low energy gamma-ray spectrum below the primary photon peak. In fact, the above equation for the FSR only works when the radiating particle is highly relativistic, $\mu_i\ll1$~\cite{Coogan:2019qpu}. Since in our case the FSR from $e^{\pm}$ dominates over $\mu^{\pm}$ and $\pi^{\pm}$ due to electron's small mass, Eq.~(\ref{eq.FSR1}) does provide a good description of the low energy spectrum.

The third term is the decay rate. The photon energy spectrum from decay is 
\begin{eqnarray}
    \frac{d N_{i,\textrm{decay}}}{dE_\gamma} &=& \frac{\Theta(E_\gamma-E_i^-) \Theta(E_i^+-E_\gamma)}{E_i^+-E_i^-}\,,\\
    E_i^\pm &=& \frac{1}{2} \left ( E_i \pm \sqrt{E_i^2 - m_i^2} \right )\,.
\end{eqnarray}
$\Theta(x)$ is the Heaviside step function. The possible photon energy is determined by the mass and energy of the  decaying particle. In this study, we only consider the two-body decay of neutral pions. The photon spectrum from pion decay is peaked at around half the pion mass. We do not include the FSR from $\mu^{\pm}$ and $\pi^{\pm}$ decays since compared to the FSR directly produced by the charged particles from the primary production, FSR from the three body decays only produce lower energy photons with more suppressed signal rates. The Hawking temperature of the PBH mass relevant for the e-ASTROGAM signal is $\sim$MeV, and the gamma-ray spectrum mainly comes from the primary photon and the electron's FSR.

Since PBH has an extended mass distribution as in Eq.~(\ref{eq:fm}), we need to integrate the flux from a single PBH with the PBH mass function to get the total flux. The observed gamma-ray flux from all PBHs is thus 
\begin{equation}\label{eq.dPhidE}
    \frac{\partial \Phi_\gamma}{\partial E_\gamma} = \bar{J}_D \frac{\Delta \Omega}{4 \pi} \int d\log m \frac{1}{m}\frac{df_\textrm{BH}}{d\log m}\frac{\partial N_{\gamma,tot}}{\partial E_\gamma \partial t}\,.
\end{equation}
$\bar{J}_D$ is the so called angular averaged $J$-factor. For Hawking radiation, $\bar{J}_D$ is the same as the decaying DM case. The $J$-factor depends on the DM halo we observe and can be calculated by integrating the DM density along the line-of-sight $\ell$ within the observation angle $\Delta\Omega$ 
\begin{equation}
    \bar{J}_D = \frac{1}{\Delta \Omega}\int_{\Delta \Omega} d\Omega \int_\textrm{LOS} d\ell \,\rho_\textrm{DM}\,.
\end{equation}
In the lower left panel of Fig.~\ref{fig:ExamplePLots}, we show the resulting gamma-ray signal from the curvature power spectra in the upper left plot. We consider signals from the galactic center with $|R|\leq 5^{\circ}$. Since the PBH abundance is highly sensitive to the curvature perturbation, a slight change of the $P_\zeta(k)$ amplitude leads to an order of magnitude difference in the gamma-ray flux. Compared to the gamma-ray signal, the peak value of $\Omega_{\rm GW}$ of the PBH GW signal only has a quadratic dependence on the $P_\zeta(k)$ amplitude. As a result, a significant change of the gamma-ray flux from varying $A$ can correspond to only a mild difference in the GW signal.

A $P_\zeta(k)$ that peaks at a higher $k_p$ (red) produces lighter PBH from collapsing smaller horizons that enter horizon at earlier times (Eq.~(\ref{eq.mpeak})). As is shown in Fig.~\ref{fig:ExamplePLots} (upper right), the red curve with larger $k_p$ peaks at a lower mass than the black curve. The production from earlier horizon re-entry also means that the PBH were formed under a higher radiation energy density.  This increases the peak value of $f_{\rm BH}$ as is shown between the same red and black curves. The larger $f_{\rm BH}$ and smaller $m$ corresponds to a higher PBH number density and generates a larger gamma-ray flux with higher $k_p$. If the $P_\zeta(k)$ is highly peaked at $k_p$ and can be approximated as a $\delta$-function, the corresponding PBH mass function will also have a narrow width. In this case , the peak of the $E^2_\gamma d\Phi_\gamma/dE_\gamma$ spectrum is mainly determined by the Hawking temperature of the peak mass
\bea
E^{\rm peak}_{\gamma}&\approx&10\,T_{\rm BH}(m^{\rm peak})\\
&\approx&1\,{\rm MeV}\,\left(\frac{5}{\gamma_{\rm eff}}\right)\,\left(\frac{k_p}{10^{15}~{\rm Mpc}^{-1}}\right)^{2}\,.\nonumber
\eea
The numerical factor $10$ is a result of the greybody factor modification on the primary photon spectrum times a small shift of the peak location from $d\Phi_\gamma/dE_\gamma$ to $E^2_\gamma d\Phi_\gamma/dE_\gamma$. The approximation explains the peak locations of the gamma-ray spectra in 
Fig.~\ref{fig:SignalBand} (upper-left), where $\gamma_{\rm eff}\approx 5$ for the range of $A_\delta$ we consider.

\begin{table*}[t]
\centering
\renewcommand{\arraystretch}{1.25}
\setlength\tabcolsep{0.75em}
\begin{tabular}{|c|c|c|c|c|c|c|c|c|}
\hline
Model & $\sigma$ & $k_p$ [Mpc$^{-1}$] & $\log_{10} A$ & $A(2\pi\sigma^2)^{-\frac{1 }{2}}$ & $f_\mathrm{BH,total}$ & $m^{\rm peak}$ [g] & $\sigma_m$ & $\gamma_{\rm eff}$\\
\hline
I & $2$ & $2\times 10^{14}$ & $-1.933$ & $2.327\times 10^{-3}$ & $1.0$ & $1.8\times10^{18}$ & $0.76$ & $3.6$\\
II & $3$ & $3\times 10^{14}$ & $-1.820$ & $2.013\times 10^{-3}$ & $1.4\times 10^{-2}$ & $6.1\times10^{17}$ & $1.0$ & $2.8$\\
III & $4$ & $3\times 10^{14}$ & $-1.737$ & $1.827\times 10^{-3}$ & $3.7\times 10^{-4}$ & $4.5\times10^{17}$ & $1.2$ & $2.0$\\
\hline
\end{tabular}
\caption{The benchmark parameters for the log-normal $P_\zeta(k)$ used in Fig.~\ref{fig:combined_contour_table}. Values of ($A$,$k_p$) were chosen to produce a minimum significance of $2\sigma$ in all experiments (although it is sometimes larger) to highlight different properties. The first three parameters are the chosen model parameters $(\sigma,k_p,\log_{10} A)$. The other five are derived. The $m^{\rm peak}, \sigma_m$ and $\gamma_{\rm eff}$ parameters are derived from a mass function fit to the log-normal distribution in Eq.~(\ref{eq:lognormalfit}). Note that due to the high sensitivity of the PBH and gamma-ray spectrum on $A$, we present our benchmark models to the fourth digit in $\log_{10} A$. The $f_{\rm BH,total}=1.0$ from Model I agrees with the bound in \cite{Carr:2017jsz} for PBH with the log-normal mass distribution. We have also verified that these models are beneath some current constraints not discussed here like the isotropic extragalactic background~\cite{Carr:2020gox} and the NuSTAR observation of the blazar TXS 0506+056~\cite{NuSTAR:2013yza}.}
\label{tab:para}
\end{table*}

\begin{figure*}[t]
    \centering
    \includegraphics[width=0.98\columnwidth]{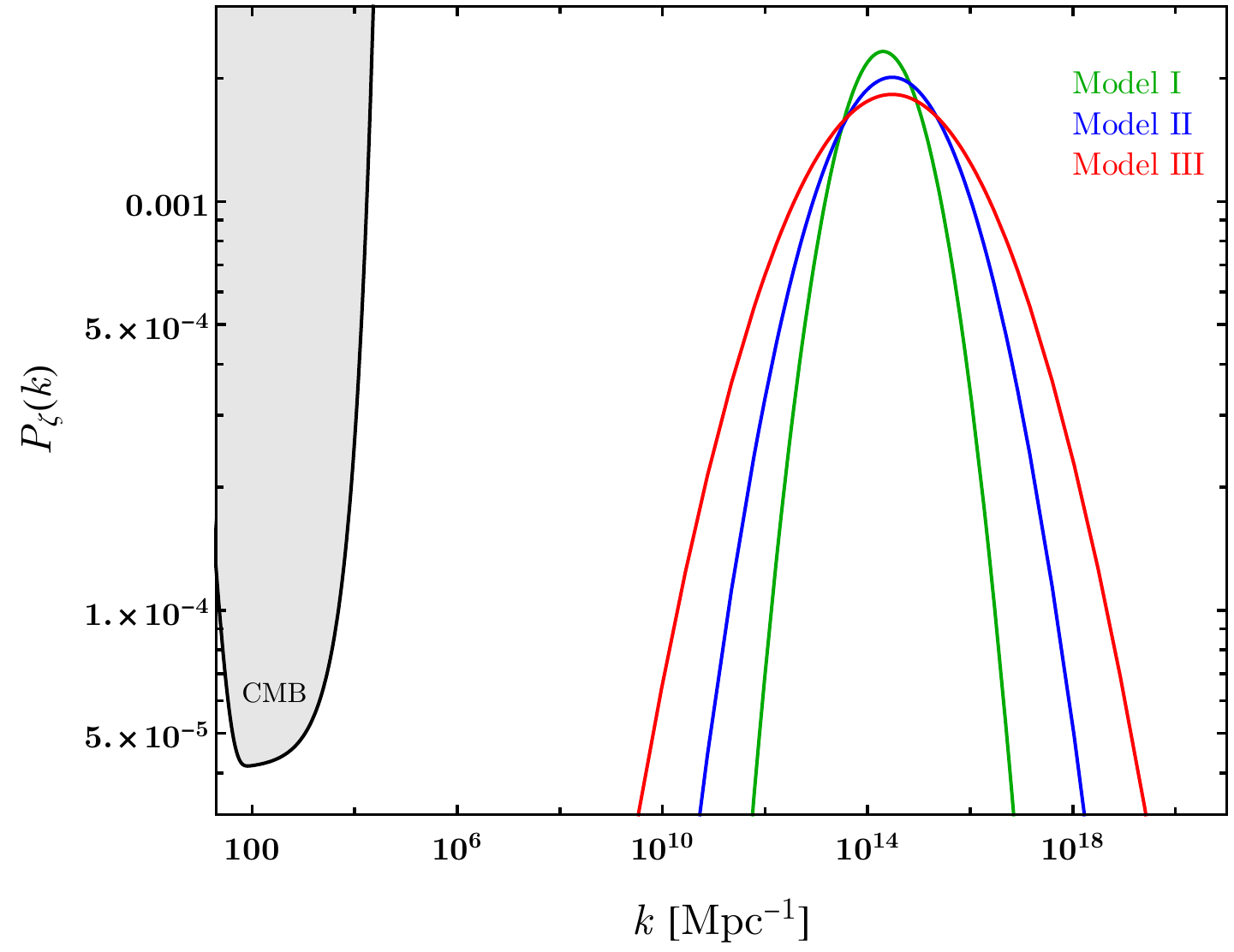}
    $\quad$
    \includegraphics[width=0.98\columnwidth]{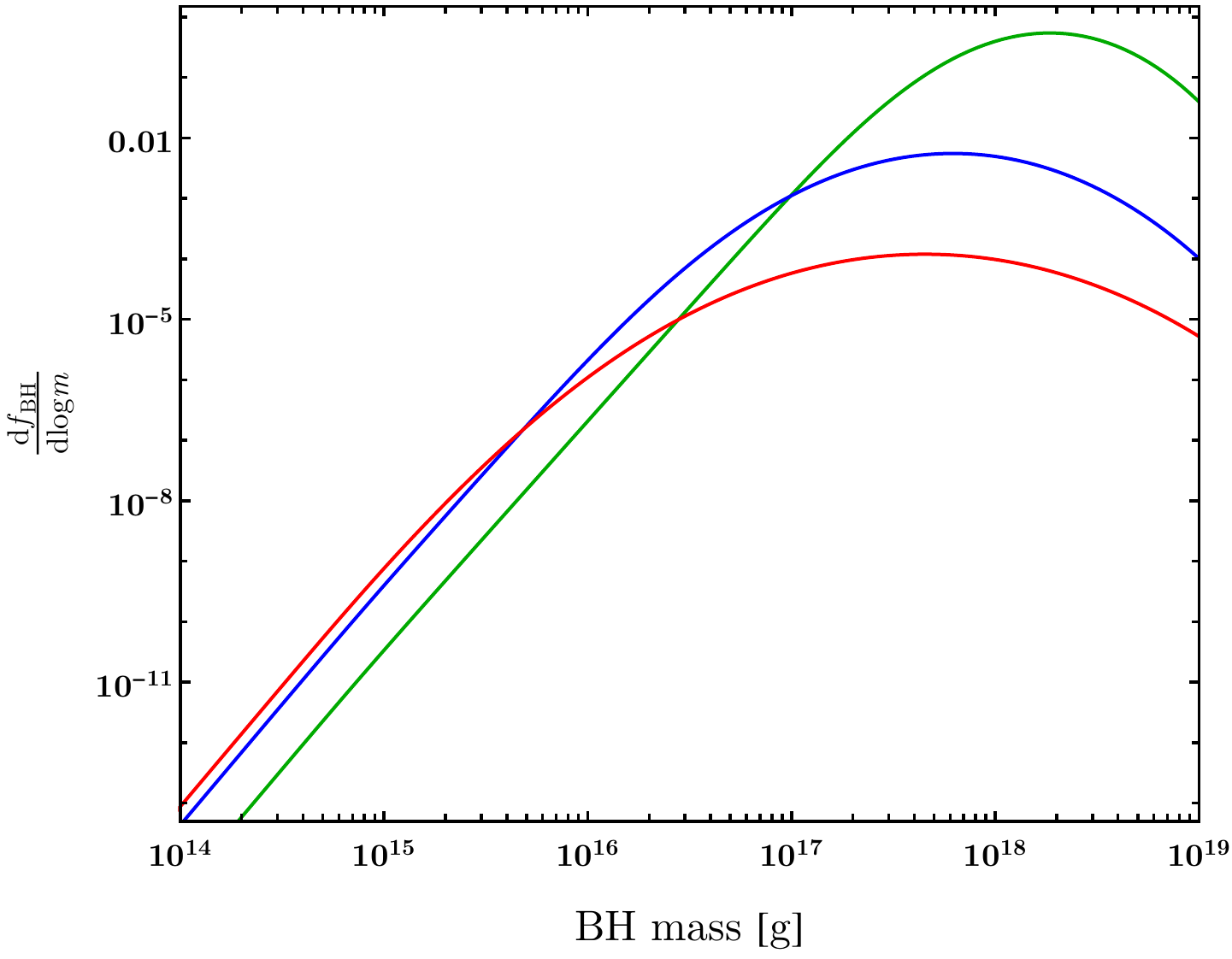}
    \\
    \includegraphics[width=0.98\columnwidth]{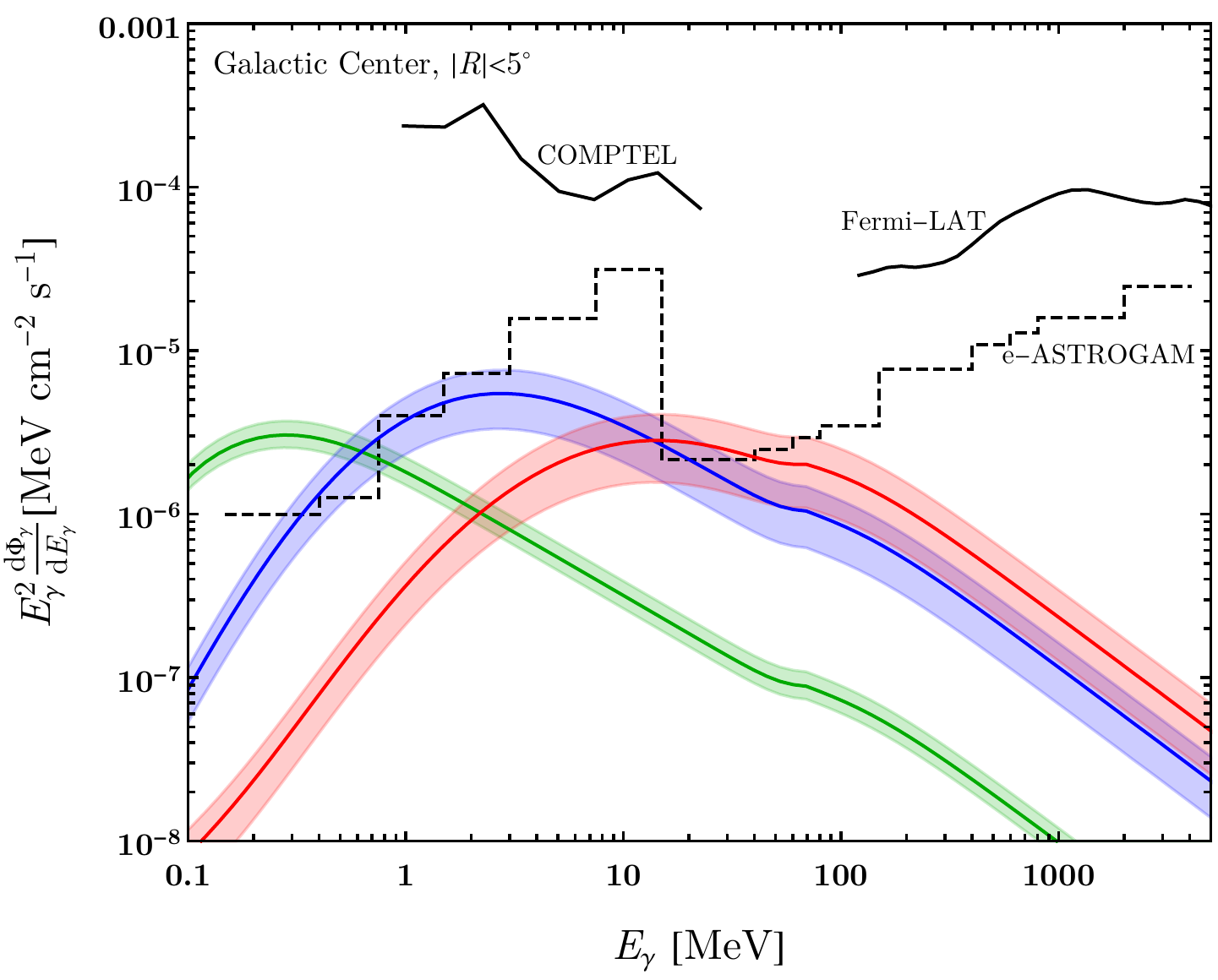}
    $\quad$
    \includegraphics[width=0.98\columnwidth]{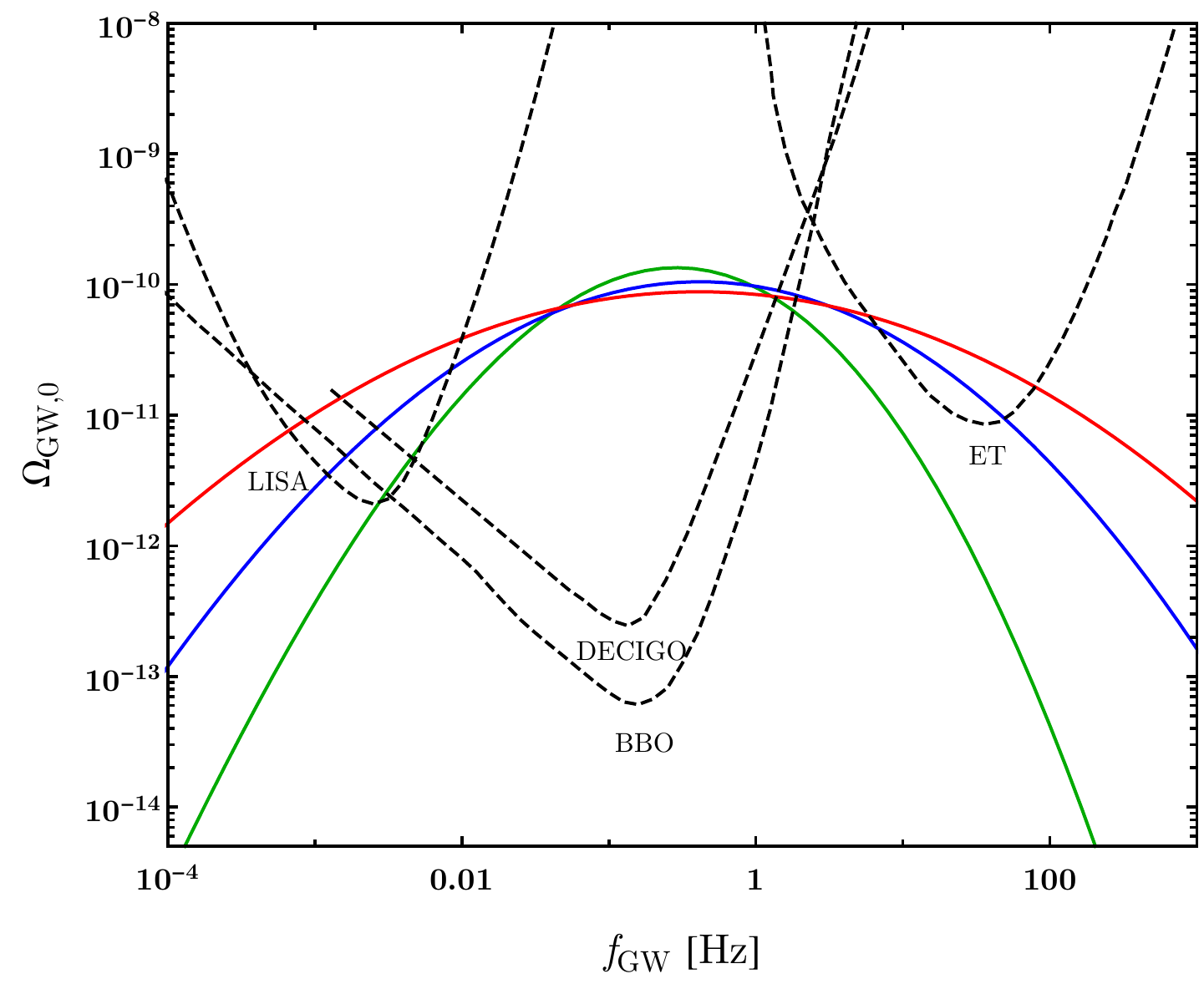}
    \caption{Power spectra and the resulting PBH and GW signals of the benchmark models in Table.~\ref{tab:para}. The grey shaded region in the top-left panel is excluded by the COBE/FIRAS measurement of CMB $\mu$-distortion~\cite{Fixsen:1996nj}. The color bands correspond to the $2\sigma$ uncertainty of the $A/(2\pi\sigma^2)^{1/2}$ measurements from the e-ASTROGAM contours (green) in Fig.~\ref{fig:combined_contour_table} with constant $(k_p,\sigma)$ for each model. The e-ASTROGAM sensitivities for each bin are at the $3\sigma$ level over the background and LISA and BBO are at the $1\sigma$ level. The DECIGO and ET sensitivity curves are taken from~\cite{Kawamura:2020pcg,Maggiore:2019uih}.} \label{fig:SignalPLots}
\end{figure*}

\section{Correlating Gamma-ray and GW signals of the PBH}\label{sec.correlate}

Given the relations between the primordial curvature perturbation Eq.~(\ref{eq:pzetaLN}), the PBH mass function Eq.~(\ref{eq:fm}), the GW Eq.~(\ref{eq.GW}), and gamma-ray signals Eq.~(\ref{eq.dPhidE}), we are in a position to correlate these two signals. We first calculate the energy of stochastic GW background from PBH that produce visible signals at e-ASTROGAM and show that the resulting GW is visible in the future GW experiments. Using three benchmark models of $P_\zeta(k)$, we further estimate the uncertainties in the fit of  curvature power spectrum from the e-ASTROGRAM and GW measurements. If we do see the gamma-ray and GW signals in the future, these uncertainties indicate how much coincidence there is for observing the same $P_\zeta(k)$ from different experiments, unless the GW and gamma-ray signals do come from PBH produced by the primordial fluctuations. 

In Fig.~\ref{fig:SignalBand}, we show the size of GW signals from the $P_\zeta(k)$ that produce PBH emitting visible gamma-ray signals in the e-ASTROGAM. To illustrate how robust the resulting GW signal can be within the DECIGO and BBO sensitivity for different shapes of $P_\zeta(k)$, we consider both the delta-function profile (upper panel) in Eq.~(\ref{eq:pzetaMC}) and the extended profile (lower panel) in Eq.~(\ref{eq:pzetaLN}) with $\sigma=4$. Each colored band corresponds to a range of the amplitude of $P_\zeta(k)$ with a fixed $k_p$. The amplitude is determined by having at least some part of the gamma-ray spectrum above the e-ASTROGAM sensitivity curve for a $3\sigma$ signal excess (dashed black) and having the whole spectrum below the existing COMPTEL and Fermi-LAT bound (solid black) and with $f_{\rm BH,\,total}\leq 1$. We will explain the derivation of the e-ASTROGAM sensitivity and the existing bounds below. As we can see, if the PBH signals show up in e-ASTROGAM, the corresponding GW signals will be well within the BBO or even ET sensitivities\footnote{The scaling behaviors of GW spectrum shape in the low frequency region are different between that from a $\delta$-function power spectrum and a narrow log-normal power spectrum. The GW from a narrow but non-zero width power spectrum scales as $f^3$ in low frequency, while the $\delta$-function power spectrum does not give this scaling~\cite{Kozaczuk:2021wcl}.}. For the extended power spectrum with $\sigma=4$, LISA can already observe the GW signal before DECIGO, BBO and ET come online. As discussed in the previous section, when rescaling the amplitude $A$, a large variation of the gamma-ray flux corresponds to a mild change to the GW signal. This means that even if the actual e-ASTROGAM performance and the background are an order of magnitude different from the estimate, there will only be a mild change to the GW signal that corresponds to the same significance in the gamma-ray excess.

In the gamma-ray plots, we obtain the Fermi-LAT bound by re-scaling the upper-limit of gamma-ray signal in Fig.~15 of~\cite{Fermi-LAT:2017opo} according to the region of interest (ROI) of Fermi observation to the ROI of our study\footnote{In \cite{Fermi-LAT:2017opo}, the ROI we use has an angular distance of $10^{\circ}$ from galactic center (GC) and blocking point sources.}. We take the COMPTEL bound from Fig.~1 of ~\cite{Essig:2013goa} with an ROI of $|b|\leq20^{\circ}, |\ell|\leq60^{\circ}$ in galactic coordinate to the ROI of our study\footnote{We assume the bound from~\cite{Essig:2013goa} is from COMPTEL observation pointing at the GC. However, we cannot confirm the data is from GC in the original COMPTEL paper~\cite{Kappadath:1998PhDT}.}. We assume the DM halo follows an NFW profile~\cite{Navarro:1996gj} with $\rho_{{\rm DM},\odot}=0.376 ~{\rm GeV}/{\rm cm}^3$, $R_{\odot}=8.122~{\rm kpc}$, $r_{200}=193~{\rm kpc}$ and $r_s=11~{\rm kpc}$~\cite{2019JCAP...10..037D}. We derive the e-ASTROGAM sensitivity using the detector and background information given in Table. III and IV of~\cite{e-ASTROGAM:2016bph}. We postpone the explanation of the sensitivity curve to the paragraph below Eq.~(\ref{eq:TS}) when  discussing the likelihood analysis for the e-ASTROGAM, LISA, and BBO. 

Suppose we do observe both the gamma-ray and GW signals, we can reconstruct $P_{\zeta}(k)$ from each of the measurements and check if results of the two completely different types of signals can reconcile with each other under the prior of the PBH formation we consider. For example, if the observed gamma-ray signal comes from DM decay or annihilation that produces a PBH-like spectrum, the GW signal will be absent. If the PBH production comes from a first order phase transition~\cite{PhysRevD.26.2681,Crawford:1982yz,Kodama:1982sf,Johnson:2011wt,Khlopov:1998nm,Liu:2021svg}, the peak frequency of GW signals relating to the e-ASTROGAM excess will be around $10^{-6}$~Hz and is below the LISA sensitivity region~\cite{Marfatia:2021hcp}. Therefore, depending on PBH's cosmic origin, it is possible that the gamma-ray observations do not have an associated GW signal in future detectors or that the GW spectrum is different from what we describe here. Even so, the measurement of both types of signals can  help to narrow down the production mechanism of the PBH. The ability of identifying the PBH and their origin, however, highly relies on the precision of  $P_{\zeta}(k)$ determination from the gamma-ray and GW data. If future experiments leave orders of magnitude uncertainties in the $P_{\zeta}(k)$ measurement, it is difficult to argue that the observed gamma-ray and GW signals have the same origin. 

To estimate the sensitivity of these measurements to the primordial perturbations, we pick three benchmark models in Table.~\ref{tab:para} as the true power spectra. The power spectrum and the resulting mass function, gamma-ray flux, and GW signals are shown in Fig.~\ref{fig:SignalPLots}. Note that while PBH are not required to constitute all of the DM in the models we are looking at, there are still regions of parameter space where 100\% PBH DM is allowed and measurable by this approach (see Model I). We conduct a likelihood study to find the $2\sigma$ uncertainties of the power spectrum parameters based on the projected sensitivity and background of the e-ASTROGAM and GW experiments. Although the PBH formation further depends on the mass parameters in Eq.~(\ref{eq:m}), we assume the parameters will be more precisely determined with better simulations of gravitational collapse. 

\begin{figure*}
    \centering
    \includegraphics[width=0.98\columnwidth]{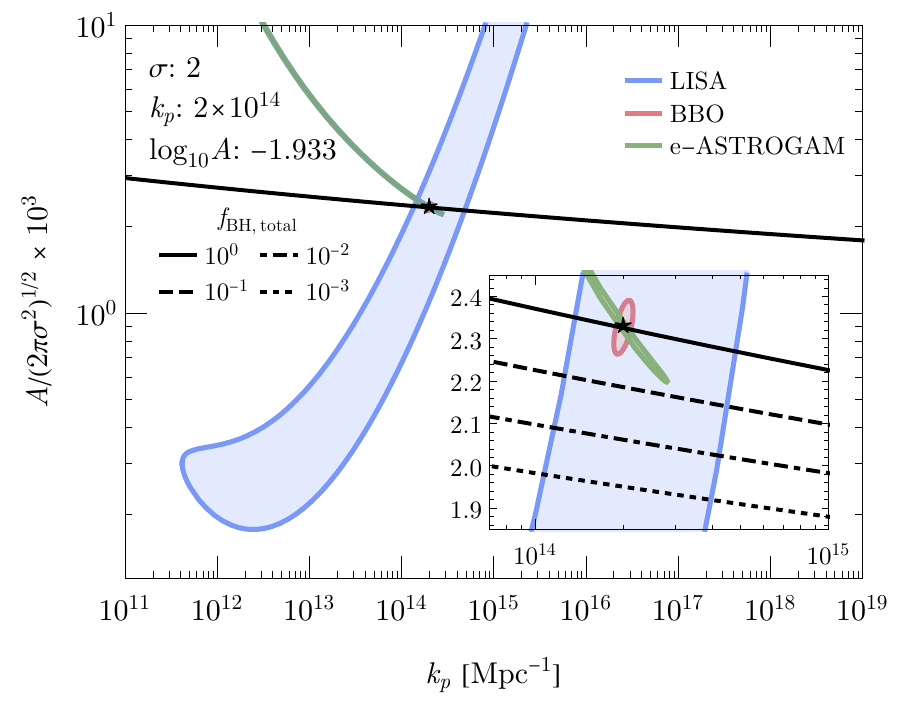}
    $\quad$
    \includegraphics[width=0.98\columnwidth]{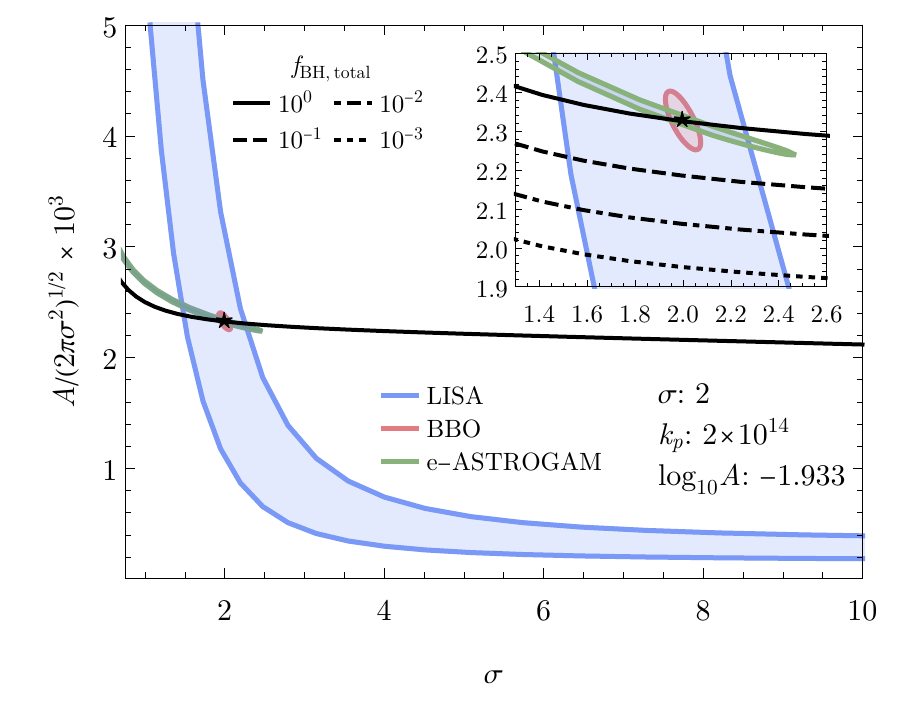}
    \\
    \includegraphics[width=0.98\columnwidth]{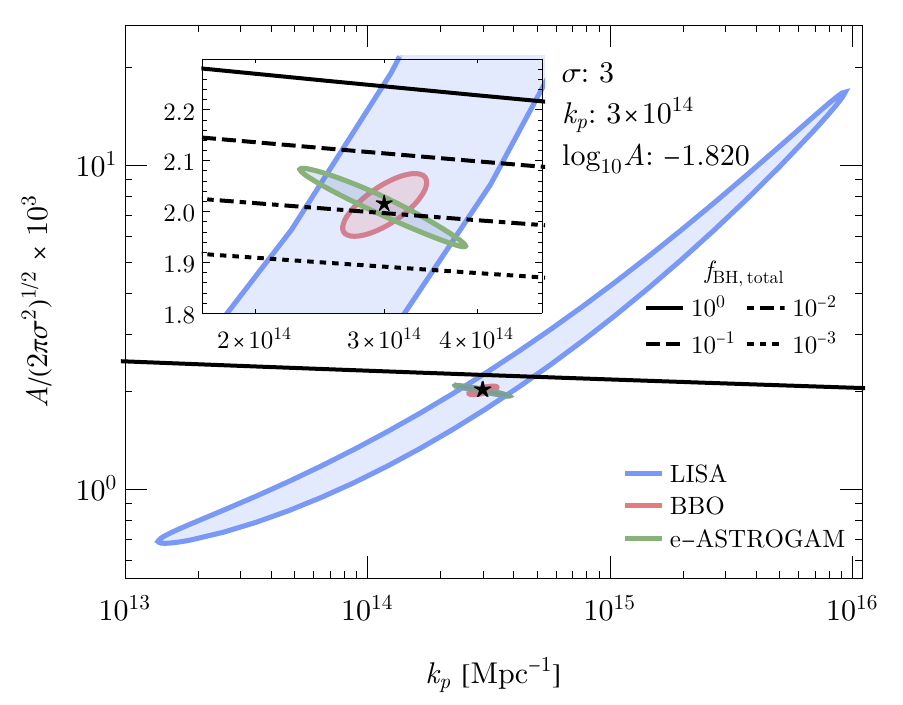}
    $\quad$
    \includegraphics[width=0.98\columnwidth]{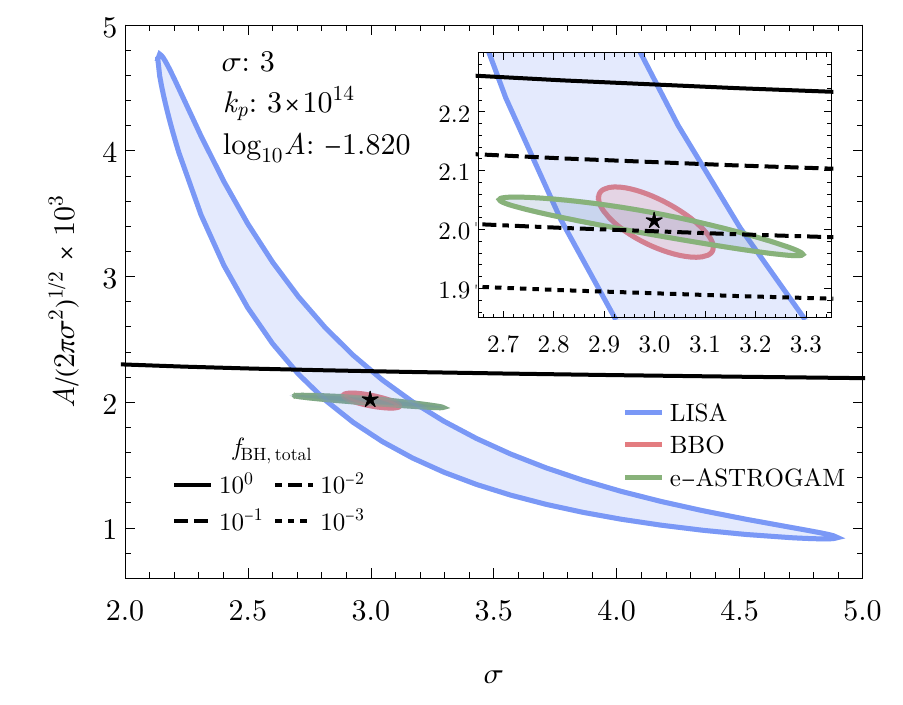}
    \\
    \includegraphics[width=0.98\columnwidth]{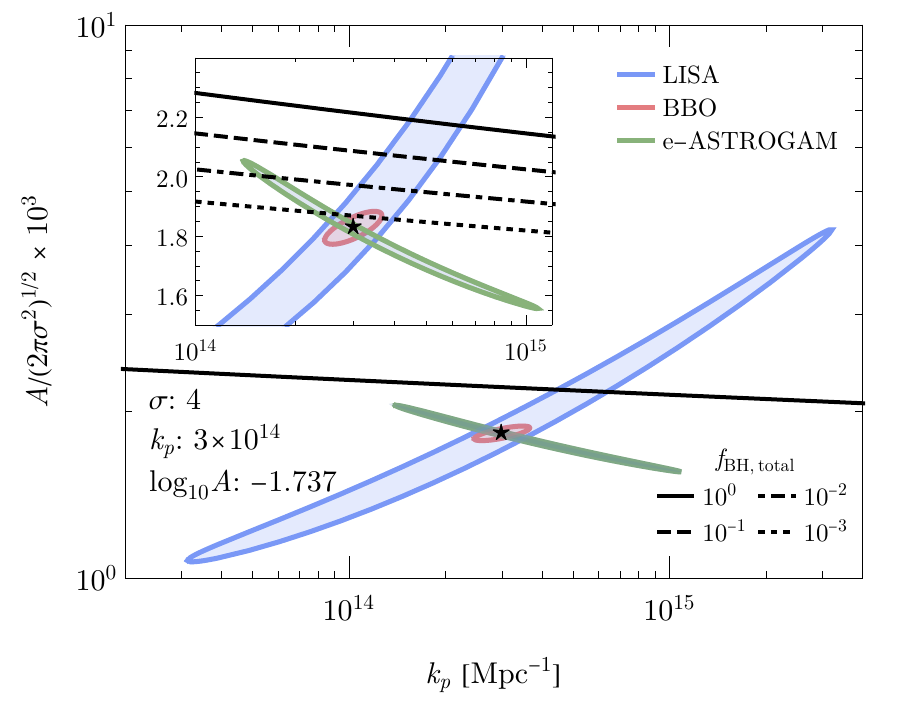}
    $\quad$
    \includegraphics[width=0.98\columnwidth]{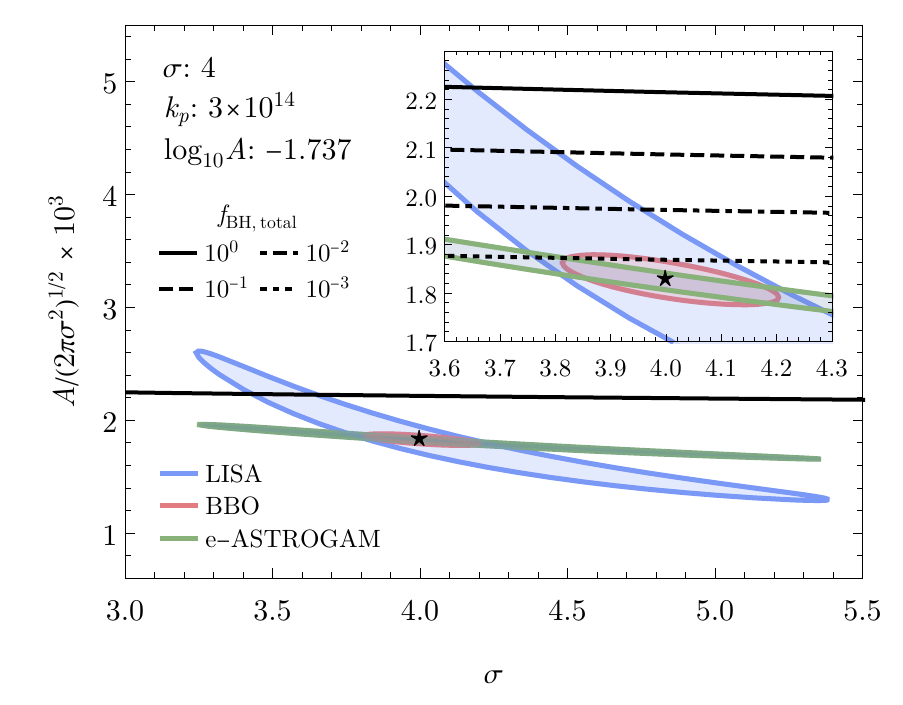}
    \caption{Differentiation of $P(k)$ spectrum models by examining GW by BBO (red), LISA (blue), and MeV photon (green) by e-ASTROGAM signals. The ``$\star$" is the true test model. The rows are arranged from top to bottom Models I/II/III. Scans were performed in the $(A(2\pi\sigma^2)^{-\frac{1 }{2}},k_p)$ ({\it left}) and $(A(2\pi\sigma^2)^{-\frac{1 }{2}},\sigma)$ ({\it right}) parameter space. The plot contours are at $2\sigma$ confidence level. The black lines (solid/dashed/dot-dashed/dotted) indicates the parameter space where $f_\mathrm{BH}= 1/10^{-1}/10^{-2}/10^{-3}$. All above the solid line is ruled out for the PBH production method used for the photon signature (green). The inset provides a zoomed in view around the true model. If approximating the PBH mass function with the log-normal distribution in Eq.~(\ref{eq:lognormalfit}), Model I/II/III has $(m^{\rm peak}\,[g],\sigma_m)=(1.8\times10^{18},0.76)/(6.1\times10^{17},1.0)/(4.5\times10^{17},1.2)$. The corresponding gamma-ray spectra are shown in Fig.~\ref{fig:SignalPLots}.}
    \label{fig:combined_contour_table}
\end{figure*}

In Fig.~\ref{fig:combined_contour_table}, we show the estimated precision of the $P_\zeta(k)$ measurements from e-ASTROGAM (green), LISA (blue), and BBO (red). We  label the parameters of the benchmark model as a black star. From the likelihood analysis that we will describe below, we draw contours either on the $(A(2\pi\sigma^2)^{-\frac{1 }{2}},k_p)$ or $(A(2\pi\sigma^2)^{-\frac{1 }{2}},\sigma)$ plane that have $2\sigma$ deviation from the signals of the benchmark model. In order to compare the different contours on a 2D plot, we fix either $\sigma$ (left) or $k_p$ (right) to the assumed numbers in Table.~\ref{tab:para} and explore the sensitivity of measuring the other two parameters. In the plots, we also show the $f_{\rm BH, total}$ curves (black) given by each set of power spectrum parameters. We do not include the mass loss as described in the end of Sec.~\ref{sec.Pk} in the $f_{\rm BH, total}$ calculation since the total DM density is mainly determined by the early universe measurements when the PBH evaporation we consider remains negligible. The allowed parameter space should be below the solid black curve for $f_{\rm BH, total}\leq 1$.

We use the detector sensitivity and astrophysical foreground studied in~\cite{e-ASTROGAM:2016bph} for the e-ASTROGAM calculation. Using their Table~3 and 4, we calculate the likelihood for the gamma-ray signal assuming that each bin is independent and follows a Poisson distribution 
\beq
\mathcal{L}_\gamma= \exp{\left(\sum_i n_i \ln{\sigma_i}-\sigma_i-\ln{n_i!}\right)}\label{eq.likeli}
\eeq
where $n_i$ is the photon count (background + model) from the assumed true source and $\sigma_i$ is the expected number of photons (once again background + model) in the bin from the test model. For the gamma-ray background, we use the estimated response values for e-ASTROGAM \cite{e-ASTROGAM:2016bph}\footnote{We use the same bins used in \cite{e-ASTROGAM:2016bph} with the minor adjustment of excluding the 10 MeV bin in the pair-production domain, as well as slightly different bin ranges in the Compton domain. These bin range changes were performed to eliminate correlation between the bins and in all cases will result in an overestimation of the background and produce conservative limits. Also note that this approach produces a different sensitivity in the pair-production domain, consistently stronger by a factor of 3 from those quoted by \cite{e-ASTROGAM:2016bph} with large backgrounds. While this does effect the sensitivity to a particular parameter set, it does not effect the results of this work and may easily be compensated by a slightly larger $A$.} utilizing the photon flux, effective area, and the stated angular resolution (the angular resolution is used to acquire the background over the entire search region). We search in a $5^\circ$ circle around the galactic center and assume a uniform background. The count rates for the background ($n_\mathrm{bk}$) and predicted source ($n_\mathrm{sig}$) can be written as
\begin{equation}
n_\mathrm{bk}=\frac{\Omega_\mathrm{region}}{\Omega_\mathrm{resol}} n_\mathrm{bk,resol}\,, \qquad n_\mathrm{sig} = \Phi_\mathrm{sig} A_\mathrm{eff}\,,
\end{equation}
where $n_\mathrm{bk,resol}$ is the counts per unit time of the background within the angular resolution patch.  $\Omega_\mathrm{region/resol}$ is the angular area of the search region and the angular resolution which was used to determine the background rate respectively. $A_\mathrm{eff}$ is the detector's effective area, and $\Phi_\mathrm{sig}$ is the predicted signal's flux. For determining the photon count, we assume an observation time of $10^8~\mathrm{s}$.

For the GW calculation, the likelihood function is written as~\cite{Caprini:2019pxz}
\begin{equation}
    \mathcal{L}_\mathrm{GW} \propto \exp{\left( -N_\mathrm{im} \sum_i \frac{1}{2} \left[ \frac{\Omega_{\mathrm{GW true},i}-\Omega_{\mathrm{GW},i}}{\sqrt{\Omega_{\mathrm{GW true},i}^2+\Omega_{\mathrm{s},i}^2}} \right]^2\right)}
    \label{eq:LogLike_GW}
\end{equation}
where $\Omega_{\mathrm{GW true},i}$ ($\Omega_{\mathrm{s},i}$) is the gravitational wave energy density in the signal (noise) for the assumed benchmark model in the $i$th frequency bin, and  $\Omega_{\mathrm{GW},i}$ is the gravitational wave energy density in the test model. $N_\mathrm{im}$ is the number of independent measurements taken for each bin and is related to observation time. This likelihood function is taken from~\cite{Caprini:2019pxz} where we have assumed the statistical averages for $\bar{D}_i$ and $\sigma_i^2$, the gravitational wave energy density and its variance respectively. We also assume that we are able to remove any foreground contamination and have an energy bin resolution of 10 equally-space logarithmic bins per decade. For the GW noise, $\Omega_\mathrm{s}$, we use the analytical approximation from~\cite{Caprini:2019pxz} for LISA. For BBO, we estimate $\Omega_\mathrm{s}$ by scaling the sensitivity curve found in \cite{Inomata:2018epa} using Eq.~(\ref{eq:LogLike_GW}). However, the exact estimate of $\Omega_s$ for BBO is unimportant due to $\Omega_\mathrm{GW} \gg \Omega_\mathrm{s}$ for most significant bins in our analysis, and the estimate is used solely for a consistent approach and to obtain an approximate shape for the noise.\footnote{The exact scaling used was $\Omega_\mathrm{s} = \Omega_\mathrm{GW} \sqrt{\frac{N_\mathrm{im}}{\Sigma^2}-1}$, where $\Omega_\mathrm{GW}$ is the sensitivity to gravitational waves taken from \cite{Inomata:2018epa}, $\Sigma$ is the confidence level of the sensitivity (we use $\Sigma=1$ as a conservative limit), and $N_\mathrm{im}$ is the number of independent measurements which we used to scale the background ($N_\mathrm{im}=94$). This equation is derived through combining Eq.~(\ref{eq:LogLike_GW}) and Eq.~(\ref{eq:TS}) and comparing with the null result.} For both LISA and BBO, we assume the number of independent measurements are comparable and use $N_\mathrm{im}=94$~\cite{Caprini:2019pxz}.

When placing estimated bounds on the model, we calculate the predicted source signal for either $\Omega_\mathrm{GW}$ (Eq.~(\ref{eq.GW})) or gamma-rays (Eq.~(\ref{eq.dPhidE})). From the assumed signals, we calculate $\mathcal{L}_\mathrm{best}$, the best-fit likelihood. We then perform a scan\footnote{During the scan, we assume that the total likelihood has a single peak at the best-fit point.} over the parameter space where we calculate the likelihood for the model and compare with the known value, $\mathcal{L}_\mathrm{best}$ from the benchmark models, by calculating
\begin{equation}
    \mathrm{TS} = - 2 \ln{\left(\frac{\mathcal{L}}{\mathcal{L}_\mathrm{best}}\right)} = \Sigma^2,
    \label{eq:TS}
\end{equation}
where $\Sigma$ is the significance of the detection~\cite{Cowan:2010js, Rolke:2004mj, Bringmann:2012vr, Fermi-LAT:2015kyq}. Here we assume the joint analysis follows a $\chi^2$ distribution. Unless stated otherwise, we use $\Sigma=2$ for all estimated constraints. The sensitivities shown in Fig.~\ref{fig:SignalBand} and \ref{fig:SignalPLots} for LISA, BBO, and e-ASTROGAM are estimated from Eq.~(\ref{eq:TS}) by setting $\Sigma$ to the desired sensitivity level, fixing $\mathcal{L}$ to the background model, and solving for the amount of signal from the true model ($\mathcal{L}_\mathrm{best}$) to satisfy the relationship. This was performed for each bin individually. 

As is shown in Fig.~\ref{fig:combined_contour_table}, the fits of the gamma-ray and GW data exhibit different degeneracy between $(A(2\pi\sigma^2)^{-\frac{1}{2}},k_p)$ or $(A(2\pi\sigma^2)^{-\frac{1 }{2}},\sigma)$. As discussed in the end of Sec.~\ref{sec.gamma}, for a fixed $\sigma$, $P_\zeta(k)$ with higher $k_p$ produces larger gamma-ray flux. Since the gamma-ray signal also grows with $A$, a similar size of the signal flux can come from models with a larger $k_p$ and a smaller $A$. This is why the green contours in the left panel of Fig.~\ref{fig:combined_contour_table} have an anti-correlation between $A(2\pi\sigma^2)^{-\frac{1 }{2}}$ and $k_p$. On the other hand, the peak $\Omega_{\rm GW}$ value of the GW signal is independent to $k_p$, and the precision of $P_\zeta(k)$ measurements from the GW are determined by the frequencies of the peak PBH signals versus the frequencies of best detection sensitivities. Since PBH that are visible in e-ASTROGAM produce GW peaks at higher frequencies than the frequencies of best sensitivities in LISA and BBO (Fig.~\ref{fig:SignalPLots}), most of the signal excess occurs on the left side of the peak. Increasing $k_p$ further enhances the peak GW frequencies (see the red vs. black curve in Fig.~\ref{fig:ExamplePLots}) thus decreasing the signal excess on the left side of the peak while increasing the excess on the right side. Because the majority of the signal within the detectors' sensitivity lies to the left of the peak, this results in an overall decrease to the excess. This explains why the red and blue contours in Fig.~\ref{fig:combined_contour_table} point to a larger  $A(2\pi\sigma^2)^{-\frac{1 }{2}}$ at higher $k_p$. When fixing $k_p$ in the right panel of the plots, increasing $\sigma$ increases both the gamma-ray and GW signals. This is why the contours point to a smaller $A(2\pi\sigma^2)^{-\frac{1}{2}}$ for larger $\sigma$ until $P_\zeta(k)$ is basically scale invariant for the $k$-modes relevant to a given search and $A(2\pi\sigma^2)^{-\frac{1}{2}}$ asymptotes to a fixed value (the LISA contour of model I). Since the sensitivities of different experiments still have different $k$-dependence, the three contours are not aligned in the same direction. 

The different parameter degeneracy from the e-ASTROGAM and LISA/BBO observations make the EM and GW signal contours to  overlap in a small region of parameter space. When assuming the observed gamma-ray and GW signals come from the PBH scenarios that we discuss, the overlap contours provide a precise determination of $P_\zeta(k)$ even if each of the contours extend along certain parameters. For example, when fixing $\sigma$ in models I and III, a combination to LISA contours in Fig.~\ref{fig:combined_contour_table} improves the $k_p$ measurement from e-ASTROGAM with orders of magnitude uncertainty to the determination of the parameter up to an $\mathcal{O}(1)$ or even $\mathcal{O}(10)\%$ level uncertainty. Once we get to see the signals from  BBO, the much smaller contour (red) will show up right on the intersection between the e-ASTROGAM and LISA contours within $\mathcal{O}(10\%)$ uncertainties around the true $(A(2\pi\sigma^2)^{-\frac{1 }{2}},k_p)$. With a wrong theory prior, the BBO result can easily miss the cross section of the e-ASTROGAM and LISA contours. A fit of the power spectrum that is right on the spot gives a strong indication that the signals do come from PBH produced by the collapse of primordial density fluctuations.

\section{Conclusion}\label{sec.conclusion}

Primordial black holes are plausible DM candidates with well-predicted Hawking radiation signals which depend only on their masses and spins. Compared to other DM candidates with a wide range of possibilities for the couplings to the Standard Model particles and hence the associated signals, this simple parametric dependence of the PBH signal gives an opportunity to uniquely  identify the source of DM using the next generation gamma-ray detectors. However, the extended PBH mass function that generally arises from a PBH production mechanism obscures this simple relation between the individual PBH properties and the gamma-ray signal. Hence a different type of observation related to the PBH is still necessary to confirm the existence of the PBH DM.

This paper demonstrates how correlating multi-messenger signals from the future gamma-ray and GW detection can provide such a desired strong evidence for the PBH DM. For this purpose, we assume that PBH emitting gamma-ray signals which are visible in the e-ASTROGAM detector come from the collapse of primordial density fluctuations. Using some examples of the curvature power spectrum, we calculate PBH gamma-ray signals that are within e-ASTROGAM's sensitivity. We require that the signal satisfy existing bounds from COMPTEL and Fermi-LAT, which set the strongest constraints on the gamma-ray flux of the $E_{\gamma}$ window we consider. We also require the produced PBH density to be within the observed DM abundance, and we also show that with the extended mass function, asteroid-mass PBH can provide $f_{\rm BH,total}=1$ (Model I). Because of the similar sensitivities, comparable results can be achieved by also considering the gamma-ray detector AMEGO.

Using these examples, we demonstrate that the same primordial fluctuations should produce GW signals which are well within the designed DECIGO, BBO, and even LISA sensitivities. By conducting a likelihood analysis with projected detector sensitivities and backgrounds, we estimate the precision with which the e-ASTROGAM, LISA and BBO measurements can determine the curvature power spectrum along these lines. In particular, as is shown in Fig.~\ref{fig:combined_contour_table}, due to the different dependence of gamma-ray and GW signals on the curvature perturbations, a combination of the e-ASTROGAM and LISA results will already/by itself significantly narrow down the power spectrum parameters around the actual values. Furthermore, observing GW signals from the BBO experiment will provide even more precise measurements of that part of the parameter space which corresponds to the overlap of the regions indicated by the e-ASTROGAM and LISA results, providing a good validation of the PBH signal 
and its production mechanism.

The observation of gamma-ray and GW signals can provide more information about the PBH than was discussed here. For example, Ref.~\cite{Agashe:2020luo} discusses the possibility of using a detailed comparison of the $J$-factors of DM signals from nearby dwarf spheroidal galaxies to distinguish various types of DM annihilation/decay processes. We can have an analogous analysis to the $J$-factors of the PBH signal that are similar to the $J$-factors of DM decay. In addition, the observation of the anisotropy of the GW signal may provide extra information about the primordial curvature fluctuations that create the PBHs ~\cite{Romano:2016dpx,Bartolo:2019zvb,Bartolo:2022pez}. Combining such different measurements further strengthens the possibility of using the EM and GW signals to shed light on PBH and their cosmic origin. We leave such a study for the future.

\section*{Acknowledgements}

The authors would like to thank Regina Caputo, Adam Coogan, Eric Cotner, Dan Hooper, Savvas Koushiappas, Alex Kusenko, Tongyan Lin, Sam McDermott, Logan Morrison, Tracy Slayter, Louie Strigari, Ethan Villarama for useful discussions. KA and JHC were supported in part by the NSF grant PHY-1914731 and by the Maryland Center for Fundamental Physics. JHC is also supported in part by JHU Joint Postdoc Fund. BD is supported in part by DOE grant DE-SC0010813. TX is supported by the Israel Science Foundation (grant No. 1112/17). YT is supported by the NSF grant PHY-2014165 and PHY-2112540.

\appendix

\section{Black Hole Mass Function Today}\label{App:massfunction}

In this Appendix, we calculate the BH mass function today from a given mass function at early times. There are two main effects, which are the expansion of the universe and the mass loss from Hawking radiation.\footnote{We ignore the effect of BH accretion here due to uncertainty of the efficiency factor. Depending on the factor, BH accretion can increase BH masses by a factor of 0.2-40 during the radiation domination \cite{Tabasi:2021cxo}.} We take a comoving frame so only the latter is relevant. The mass loss rate for a black hole mass $m$ can be calculated by integrating the Hawking radiation spectrum in Eq.~(\ref{eq:HRprimary}),
\bea
\dot{m}(m) &=& - \frac{\mathcal{G} g_{\star H}(T_\BH(m)) \mPl^4}{30720 \pi \, m^2}\\
&=& - f(m) \frac{\mPl^4}{m^2},
\eea
where $\dot{m}=\frac{\dt m}{\dt t}$, $\mathcal{G}$ is from the greybody factor, $g_{\star H}(T_\BH)$ counts all the particle species for Hawking radiation at the temperature $T_\BH (m)$. A more detailed description can be found in~\cite{Hooper:2019gtx}. We took the numerical values of $f(m)$ from BlackHawk \cite{Arbey:2019mbc, Arbey:2021mbl}. The mass spectrum today can be obtained by solving the partial differential equation,
\beq\label{eq:nBHPDE}
{\dd t} n_\BH(t,m) + {\dd m} \left( \dot{m}(m) \, n_\BH(t,m) \right) = 0,
\eeq
where $n_\BH(t,m)$ is the number density of PBH per unit mass. $f(m)$ has mild dependence on $m$, so it can be considered as a constant $f_0$. The analytic solution for Eq.~\ref{eq:nBHPDE} with this approximation is
\beq 
n_\BH(t,m) = \frac{m^2 n_{\BH}(t_i,(m^3+3f_0 \mPl^4 t)^{1/3})}{(m^3+3f_0 \mPl^4 t)^{2/3}},
\eeq
where $t_i$ is the formation time. We also assume $t_i \ll t$. We have checked numerically that using $f(m)=f(0.7372 \, m_\textrm{evp}) \approx 1.895 \times 10^{-3}$ gives results with $\mathcal{O}(1\%)$ errors for interesting parameters, where $m_\textrm{evp}$ is the black hole mass evaporating today.

\section{Rescaling the gamma-ray bounds}
We re-scale gamma-ray bounds from COMPTEL~\cite{Essig:2013goa} and Fermi-LAT~\cite{Fermi-LAT:2017opo} to the ROI of this study in the following way: we use gamma-ray bounds in \cite{Essig:2013goa,Fermi-LAT:2017opo} and the original ROI to obtain the upper limit on PBH number density, then we calculate the gamma-ray flux with our ROI of $|R|<5^{\circ}$ from the PBH number density saturating the existing bounds. For the COMPTEL bound, the allowed signal flux is obtained from 
\beq
\left. \frac{d\Phi}{dE} \right|_{|R|<5^{\circ}} \! =\frac{(J_{D} \: \Delta\Omega)|_{|R|<5^{\circ}}}{(J_{D} \: \Delta\Omega)|_{|b|<20^{\circ},|\ell|<60^{\circ}}} \! \left. \frac{d\Phi}{dE} \right|_{|b|<20^{\circ},|\ell|<60^{\circ}}
\eeq
For Fermi-LAT, we have
\beq
\left. \frac{d\Phi}{dE} \right|_{|R|<5^{\circ}} \! =\frac{(J_{D} \: \Delta\Omega)|_{|R|<5^{\circ}}}{(J_{D} \: \Delta\Omega)|_{|R|<10^{\circ}}} \! \left. \frac{d\Phi}{dE} \right|_{|R|<10^{\circ}}\,.
\eeq
\begin{figure*}[t]
    \centering
    \includegraphics[width=0.98\columnwidth]{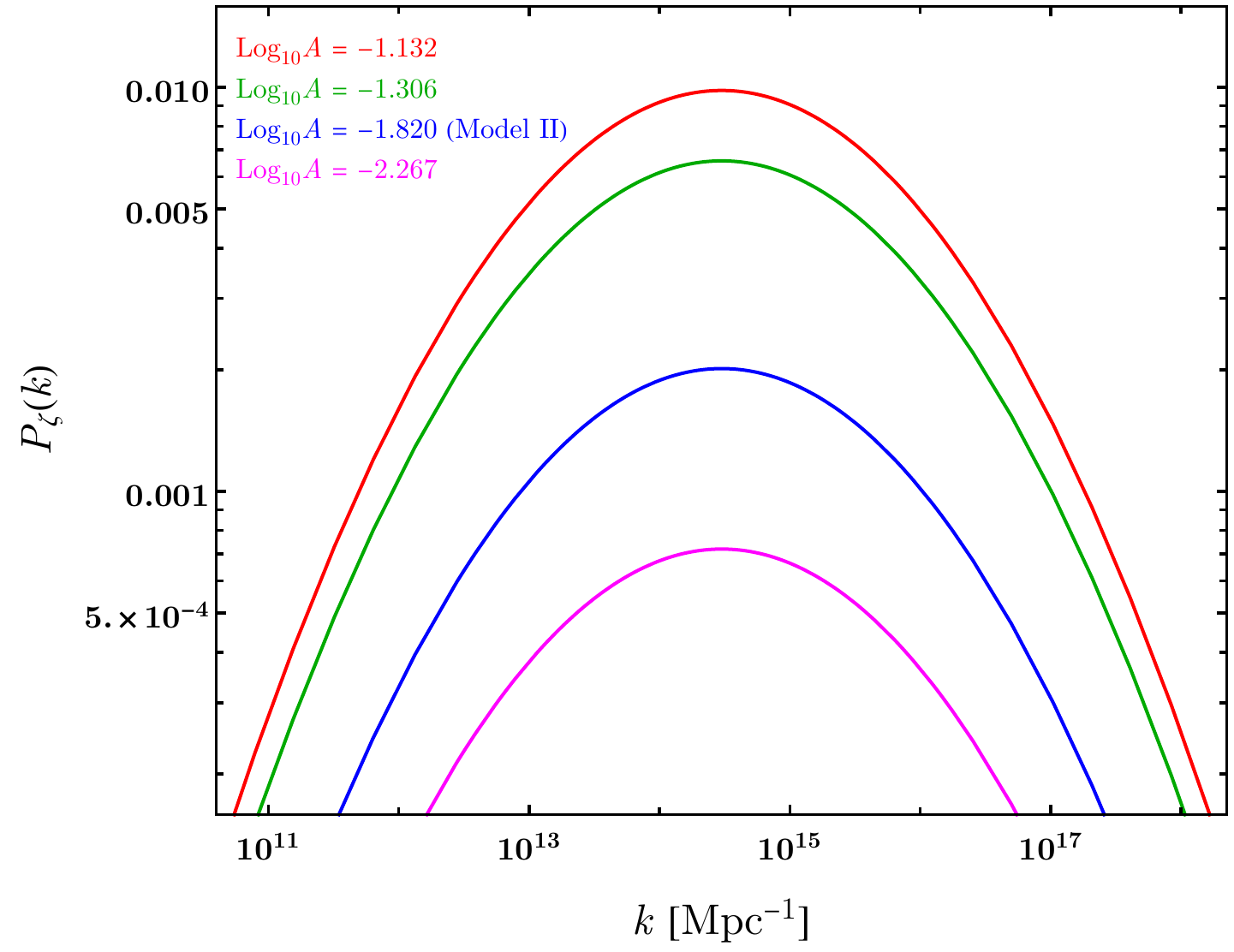}
    $\quad$
    \includegraphics[width=0.98\columnwidth]{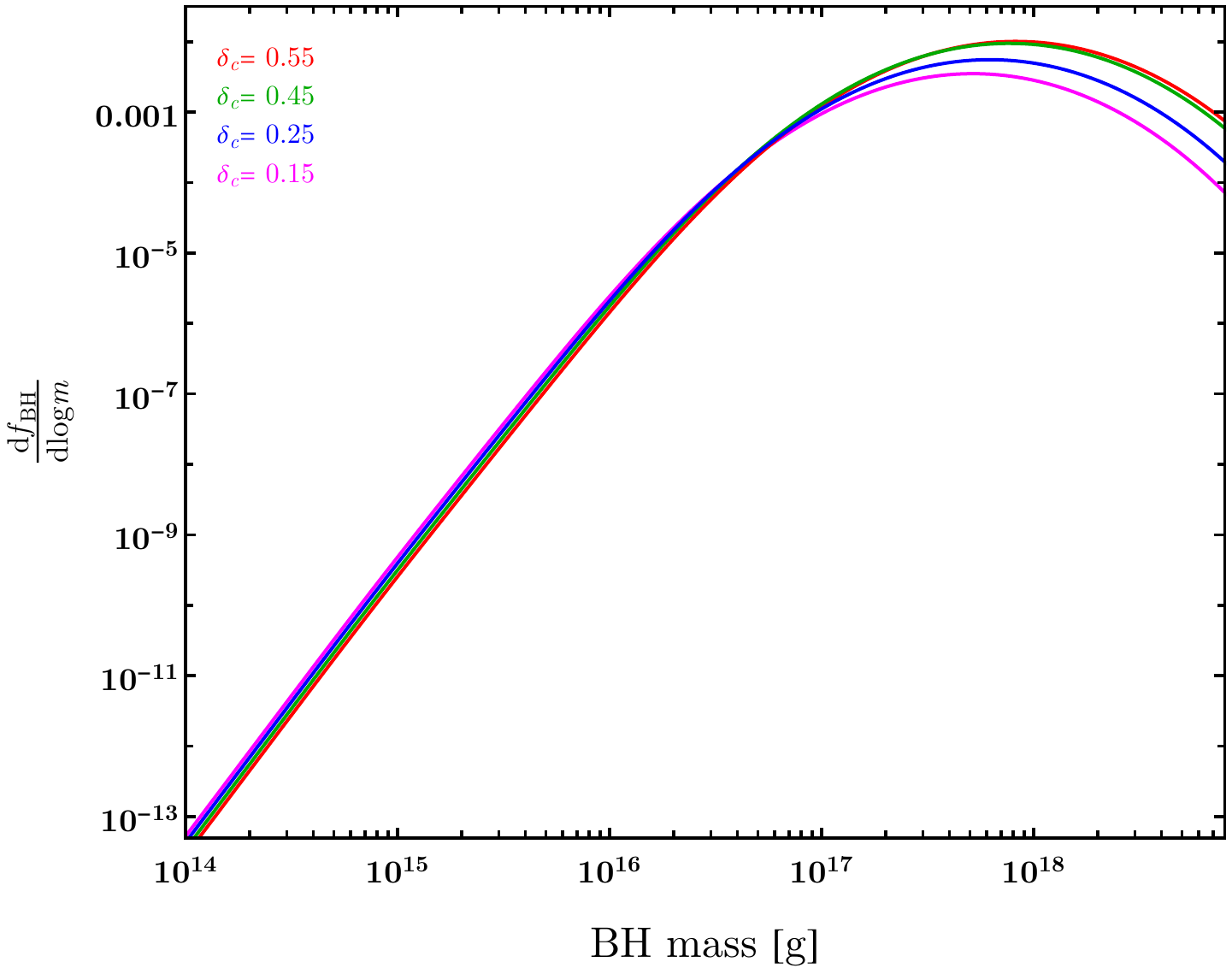}
    \\
    \includegraphics[width=0.98\columnwidth]{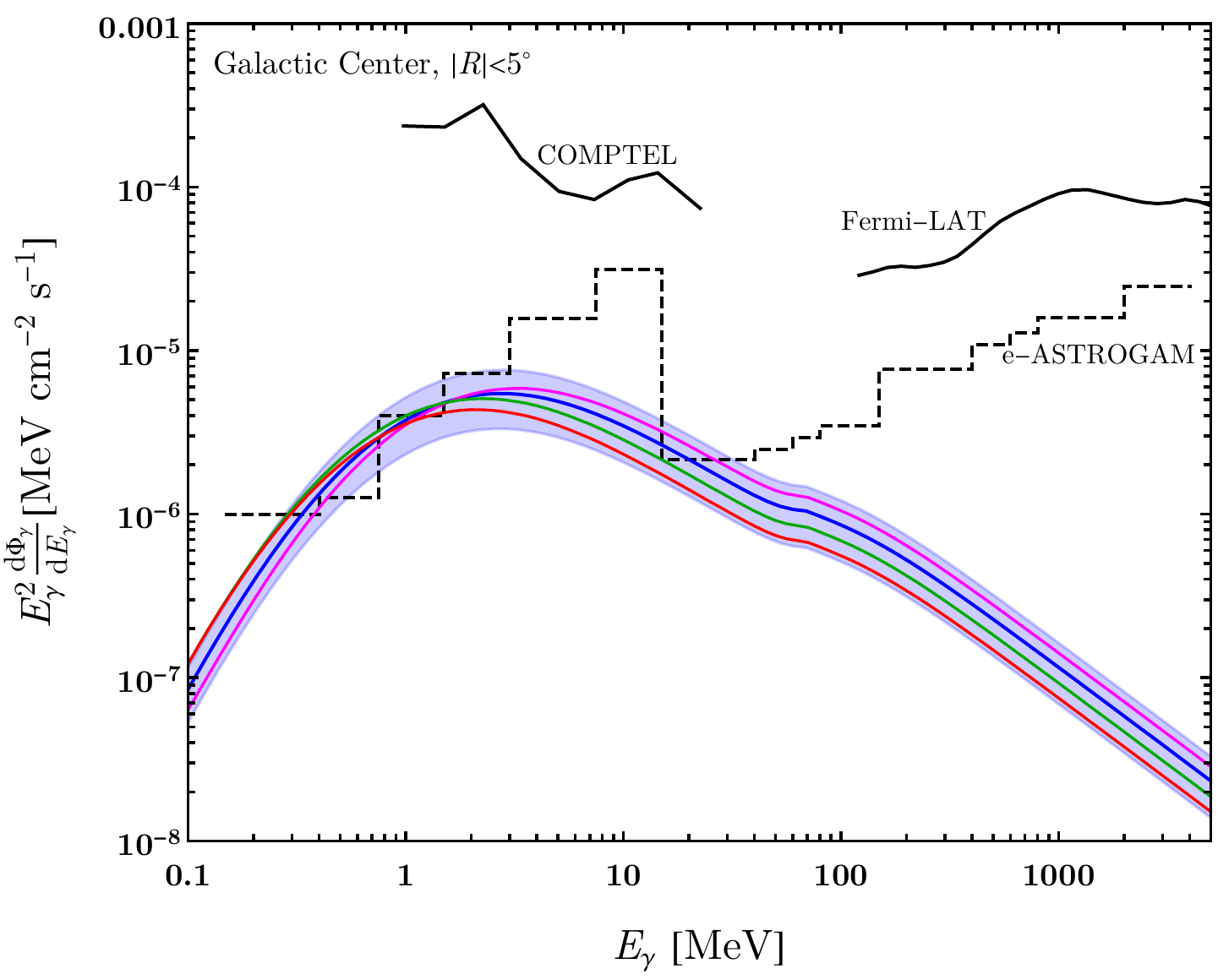}
    $\quad$
    \includegraphics[width=0.98\columnwidth]{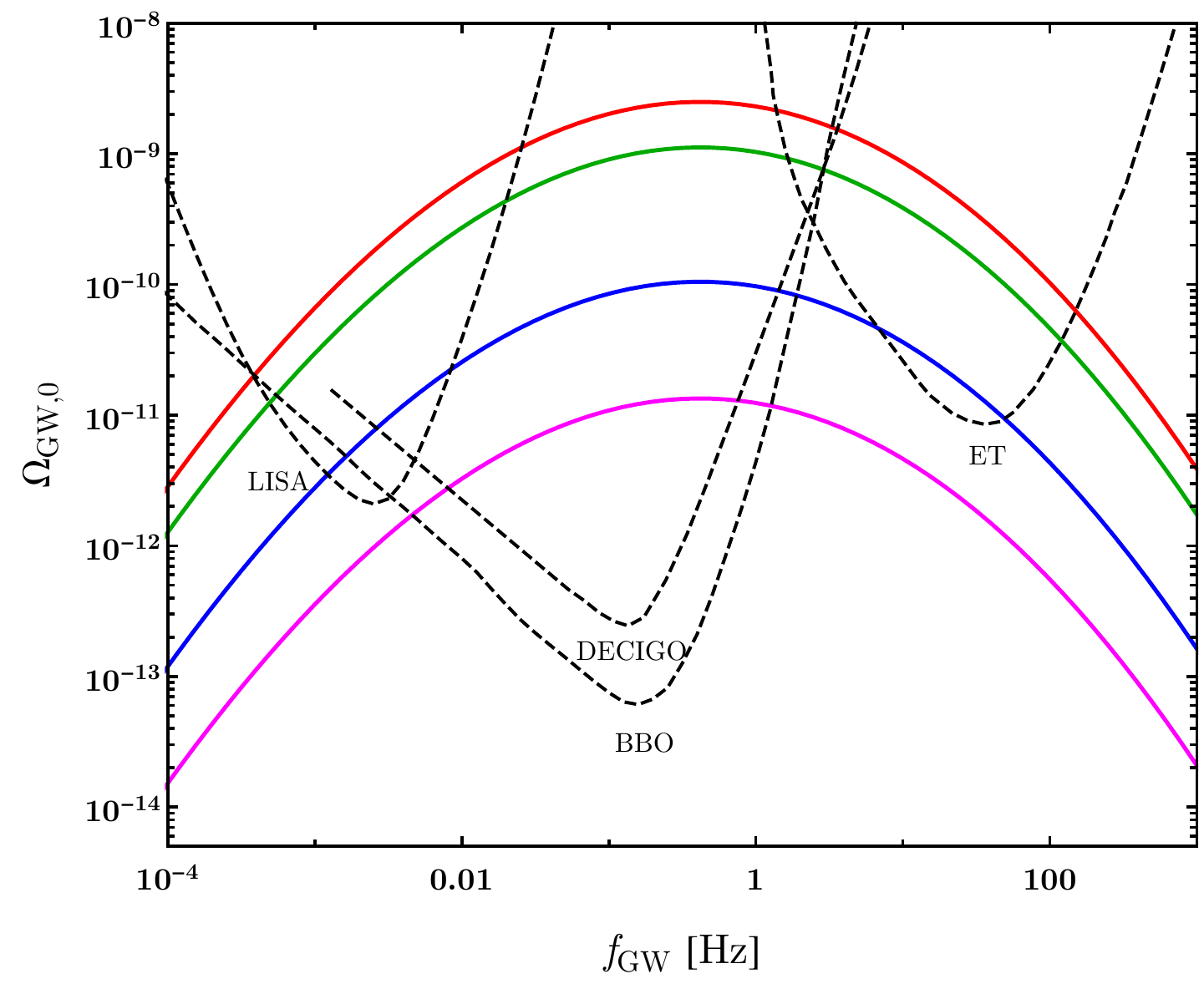}
    \caption{Example plots with difference threshold value $\delta_c$. Top-left: examples of curvature power spectrum following the log-normal distribution in Eq.~(\ref{eq:pzetaLN}) with different amplitude $A$ values. The $\sigma=3$ and $k_p=3\times10^{14}~{\rm Mpc}^{-1}$ are the same as Model II in Table.~\ref{tab:para}. Top-right: the resulting PBH mass spectra. Bottom-left: the galactic center gamma-ray flux from Hawking radiation in the ROI of $5^{\circ}$. Bottom-right: GW signal spectrum from the curvature power spectrum.} \label{fig:DeltacPlot}
\end{figure*}

\section{PBH formation threshold $\delta_c$}\label{app.deltac}

To give an idea on how the uncertainty in $\delta_c$ changes the PBH signal, we perform the same calculation of the PBH mass function described in Sec.~\ref{sec.Pk} but with different $\delta_c$ values. Note that such a calculation may not be consistent with the choice of $(K,\gamma,\delta_c)$ and the window function used in~\cite{Young:2020xmk} and should only be treated as an illustration of the $\delta_c$-dependence in the estimate.

Ref.~\cite{Musco:2018rwt,Escriva:2019phb,Musco:2020jjb} find a range $0.4\lsim\delta_c\lsim 2/3$ for different initial curvature spectra and top-hat profiles. After applying the factor of $2.17$ difference obtained in~\cite{Young:2020xmk} for a Gaussian smoothing function, the range becomes $0.2\lsim\delta_c\lsim 0.3$. For a conservative assumption of the possible variation of the $\delta_c$ values, we blindly choose four different threshold values,  $\delta_c=0.15,\,0.25,\,0.45,\,0.55$ in the calculation. We plot the gamma-ray and GW signals in Fig.~\ref{fig:DeltacPlot} for Model II in Table.~\ref{tab:para} with a normalization in $A$ that gives the similar excess in the e-ASTROGAM detection. Compared to the original calculation with $(K,\delta_c,\gamma)=(10,0.25,0.36)$ (blue), we produce similar $df_{\rm BH}/dm$ and gamma-ray signals for the other choices of $\delta_c$ by re-scaling the amplitude of $P_{\zeta}(k)$ with a factor between $0.36-4.9$ (from magenta to red) that roughly equals $(\delta_{c,{\rm new}}/\delta_{c,{\rm original}})^2$. The resulting $\Omega_{\rm GW}$ changes by a factor between $0.13-24$ due to its $A^2$-dependence. Given that the GW signals we consider in the main text are well-above the BBO sensitivities and below the existing bound ($\Omega_{\rm GW}\lsim 1.7\times10^{-7}\,(2\sigma)$ at $f=20-86$~Hz from LIGO~\cite{LIGOScientific:2016jlg}), the change of the GW signal will not modify our conclusion of seeing both the gamma-ray and GW signals from PBH.

\bibliographystyle{utphys}
\bibliography{bibliography.bib}

\end{document}